\documentclass[aps,prd,twocolumn,showpacs,superscriptaddress,nofootinbib,preprintnumbers]{revtex4-2}

\usepackage{style}

\begin{document}

\preprint{FERMILAB-PUB-24-0752-T}

\title{Observable CMB $B$-modes from Cosmological Phase Transitions}

\author{Kylar Greene\,\orcidlink{0000-0002-2711-7191}}
\email{kygreene@unm.edu}
\affiliation{Department of Physics and Astronomy, University of New Mexico
Albuquerque, New Mexico 87131}
\affiliation{Theoretical Physics Division, Fermi National Accelerator Laboratory, Batavia, Illinois 60510}

\author{Aurora Ireland\,\orcidlink{0000-0001-5393-0971}} 
\email{anireland@stanford.edu}
\affiliation{Stanford Institute for Theoretical Physics, Department of Physics, Stanford University, Stanford, CA 94305}

\author{Gordan Krnjaic\,\orcidlink{0000-0001-7420-9577}} 
\email{krnjaicg@uchicago.edu}
\affiliation{Theoretical Physics Division, Fermi National Accelerator Laboratory, Batavia, Illinois 60510}
\affiliation{Kavli Institute for Cosmological Physics, University of Chicago, Chicago, IL 60637}
\affiliation{Department of Astronomy and Astrophysics, University of Chicago, Chicago, IL 60637}

\author{Yuhsin Tsai\,\orcidlink{0000-0001-7847-225X}}
\email{ytsai3@nd.edu}
\affiliation{Department of Physics and Astronomy, University of Notre Dame, South Bend, IN 46556}

\date{\today}

\begin{abstract}
A $B$-mode polarization signal in the cosmic microwave background (CMB) is widely regarded as smoking gun evidence for gravitational waves produced during inflation. Here, we demonstrate that tensor perturbations sourced during non-inflationary epochs can yield non-negligible B-mode signals, which can in principle complicate the interpretation of future observational data. As a case study, we consider tensor perturbations sourced in the bubble collision stage of a first-order cosmological phase transition occurring in a secluded dark sector. Although phase transitions arise from causal sub-horizon physics, they nevertheless exhibit a white noise power spectrum on super-horizon scales. Power is suppressed on the large scales relevant for CMB $B$-mode polarization, but it is not necessarily negligible. We show that for appropriately chosen phase transition parameters, the maximal $B$-mode amplitude can compete with inflationary predictions that can be tested with current and future experiments. These scenarios can be differentiated by performing measurements on multiple angular scales, since the phase transition signal predicts peak power on smaller scales.
\end{abstract}
\bigskip
\maketitle

\section{Introduction}\label{sec:intro}

Cosmological inflation is a compelling framework for dynamically solving the horizon and flatness problems while also generating the density perturbations observed in our universe today \cite{Baumann:2009ds,Ellis:2023wic}. Inflationary models also generically predict nearly scale-invariant tensor perturbations, which induce a characteristic $B$-mode polarization signal in the cosmic microwave background (CMB) \cite{Kosowsky:1994,Kamionkowski:1996zd,Seljak:1996gy,Kamionkowski:1996ks,Zaldarriaga:1996xe,Seljak:1996ti,Hu:1997hv,Dodelson:2003bv,Smith:2005mm,Mortonson:2009qv,Kamionkowski:2015yta,Guzzetti:2016mkm}. It is widely accepted that observing $B$-modes above known astrophysical foregrounds would constitute ``smoking gun" evidence for inflation \cite{Weinberg:2003ur,Flauger:2007es,Baumann:2014cja,BICEP2:2014owc}.  

In this \textit{Letter}, we present a counterexample of post-inflationary $B$-modes that can rival the signal strength of testable inflationary predictions. Indeed, {\it any} source of large-scale,  coherent tensor perturbations produced before reionization can contribute to $B$-mode signals.\footnote{See also Ref.~\cite{Geller:2021obo} for $B$-mode signals from resonant particle production near reionization.} Our representative scenario is a late-time, strongly first-order phase transition occurring in a secluded dark sector, which produces gravitational waves (GWs) through bubble collisions.\footnote{First-order cosmological phase transitions also source GWs through sound waves and turbulence. We consider a supercooled transition for which these contributions are subdominant.} The key difference with respect to inflationary signals is that the tensor power spectrum from sub-horizon sources is not (nearly) scale invariant, but rather white noise on super-horizon scales. As a result, the $B$-mode signal has a distinct spectral shape, with more power on smaller scales as compared with the inflationary prediction. Accurately measuring $B$-modes on different angular scales can then distinguish between these sources.

The GW signal from a cosmological phase transition has been extensively studied in pioneering earlier works~\cite{Kosowsky:1992rz,Kamionkowski:1993fg,Kosowsky:2001xp,Apreda:2001us,Nicolis:2003tg,Grojean:2006bp,Huber:2008hg,Kahniashvili:2008pf,Kahniashvili:2009mf,Caprini:2009fx,Espinosa:2010hh,Hindmarsh:2013xza,Caprini:2015zlo,Caprini:2015tfa,Kisslinger:2015hua,Hindmarsh:2015qta,Schwaller:2015tja,Dev:2016feu,Jinno:2017fby,Hindmarsh:2017gnf,Weir:2017wfa,Brdar:2018num,Caprini:2018mtu,Mazumdar:2018dfl,Cutting:2018tjt,Geller:2018mwu,Alanne:2019bsm,Hindmarsh:2019phv,Schmitz:2020syl,Hindmarsh:2020hop,Ellis:2020awk,Cutting:2020nla,Athron:2023xlk,Caprini:2024gyk}, including analytical estimates of the GW spectrum~\cite{Caprini:2007,Jinno:2016,Caprini:2018}. Recently, it has also been shown that the white noise scalar perturbations from bubble nucleation can affect CMB temperature anisotropy measurements \cite{Elor:2023xbz}. To our knowledge, however, the CMB $B$-mode signal from a first-order phase transition has not been calculated before.  
 
This {\it Letter} is organized as follows: Sec.~\ref{sec:ClBB} develops the formalism for calculating the $B$-mode polarization signal; Sec.~\ref{sec:PTfromPT} reviews the tensor power spectrum from a phase transition; Sec.~\ref{sec:numerical} presents our numerical results; Sec.~\ref{sec:OmegaGW} discusses the complimentary GW signal; and Sec.~\ref{sec:discussion} offers some concluding remarks and future directions. 


\section{B-mode Polarization Signal}\label{sec:ClBB}

Temperature anisotropies in the CMB arise from scalar, vector, and tensor perturbations. When photons scatter with free electrons, quadrupole anisotropies in the temperature distribution are converted to polarization of the
scattered photons. The angular spectrum of $B$-mode polarization can be written
\begin{equation}\label{eq:ClBB}
    C_\ell^{BB} = 36 \pi \int_0^\infty \frac{dk}{k} \mathcal{P}_h(k) \mathcal{F}_\ell(k)^2 \,,
\end{equation}
where $\mathcal{P}_h(k)$ is the dimensionless tensor power spectrum evaluated at the initial time and
\be\label{eq:Fl}
    \mathcal{F}_\ell (k) &=& \int_0^{\tau_0} d\tau \, V(\tau_0,\tau) \, {\cal S}_\ell(k, \tau_0, \tau)      \int_0^\tau d\tau_1 \, V(\tau,\tau_1)  \nonumber
    \\ 
    && ~~~\times
  \int_{\tau_1}^{\tau} d\tau_2  \, \frac{j_2[k(\tau - \tau_2)]}{k^2 (\tau - \tau_2)^2} 
  \brac{\partial \mathcal{T}(\tau_2,k)}{\partial \tau_2}
  ,~~~~~ 
\ee
where $\tau$ is the conformal time, $\tau_0$ is its value today, $V(\tau_1, \tau_2)$ is the visibility function, $\mathcal{T}(\tau,\vec{k})$ is the transfer function, $j_n(x)$ is a spherical Bessel function of the first kind, and 
\be
\label{eq:Sfunc}
{\cal S}_\ell(k, \tau_0, \!\tau) \equiv  \frac{\ell \!+\! 2}{2 \ell \! +\! 1} j_{\ell - 1} \!\left[ k(\tau_0 \!-\! \tau) \right] \!- \!\frac{\ell \!-\! 1}{2\ell \!+\! 1} j_{\ell + 1}\! \left[ k (\tau_0 \!-\! \tau) \right] \! .~~~~~~ 
\ee
See Sec.~I of the Supplemental Materials for a derivation and discussion. 

This formula presumes a decomposition of the tensor perturbation $h_{ij}(\tau, \vec{k}) = h_{ij}^{\rm ini}(\vec{k}) \mathcal{T} (\tau, \vec{k})$, where $h^{\rm ini}_{ij}(\vec{k})$ is the initial perturbation amplitude. This decomposition is useful because it separates the statistical correlations of the initial amplitudes from the deterministic effect of the modes' subsequent evolution, as captured by the transfer function. The statistical properties of the initial perturbations are encoded in the (dimensionless) tensor power spectrum $\mathcal{P}_h(k)$, 
\be
\label{eq:Ph}
    \left\langle \! h^{\rm ini}_{ij}(\vec{k}) \, h^{\rm ini}_{i'j'}(\vec{k}')^*  \!\right\rangle \! = \! \frac{\delta_{i i'}\delta_{j j'}}{2} \frac{2\pi^2}{k^3}  \mathcal{P}_h(k) (2\pi)^3 \delta(\vec{k} - \vec{k}'). ~~~~
\ee 
The metric perturbation $h_{ij}(\tau,\vec{k})$ evolves according to
\begin{equation}\label{eq:fullwaveeq}
    h_{ij}'' + 2 \mathcal{H} h_{ij}' + k^2 h_{ij} = 8 \pi G a^2 \Pi_{ij} \,,
\end{equation}
where $\mathcal{H} = a'/a$ is the conformal Hubble rate, primes denote derivatives with respect to $\tau$, and the source $\Pi_{ij}(\tau, \vec{k})$ is the Fourier-transformed anisotropic stress. Solving this equation for $h_{ij}$ and evaluating at the end of the phase transition will give us the initial amplitude $h_{ij}^{\rm ini}$. Afterwards, modes evolve according to the transfer function, which satisfies Eq.~(\ref{eq:fullwaveeq}) without the source term. See Ref.~\cite{Kite:2021} for further detail on the transfer function.

\section{Tensor Power Spectrum from Bubble Collisions}\label{sec:PTfromPT}

First-order cosmological phase transitions (PTs) proceed through bubble nucleation and source tensor perturbations in three stages: the bubble collision stage~\cite{Hawking:1982ga,Kosowsky:1992vn}, the acoustic stage~\cite{Hindmarsh:2013xza,Hindmarsh:2016lnk,Hindmarsh:2019phv,RoperPol:2023dzg}, and the turbulent stage~\cite{Caprini:2006jb,Kahniashvili:2008pe,Caprini:2009yp,Auclair:2022jod}. The relative contribution from each of these phases depends largely on the PT strength, as parameterized by $\alpha$, as well as the bubble wall velocity $v_w$. 

As a case study, we consider a strongly-supercooled PT with ``runaway'' bubble wall $v_w \rightarrow 1$, for which the dominant contribution comes from the bubble collision stage~\cite{Ellis:2019}. Because such a scenario requires negligible plasma friction, it is difficult to realize within the Standard Model. We therefore assume the PT occurs in a thermally-decoupled dark sector. 

In what follows, we focus exclusively on the contribution from bubble collisions. Including sound wave contributions can further increase the signal, particularly for the benchmark points we later consider. However, because this source is more difficult to compute semi-analytically and because the details depend on model-dependent properties of the dark sector fluid, we leave this to future work. Our calculation should thus be viewed as a conservative proof-of-principle, rather than the definitive prediction of a particular model.

Following Refs.~\cite{Caprini:2007,Jinno:2016}, during the PT we solve Eq.~(\ref{eq:fullwaveeq}) with source $\Pi_{ij}(\tau,\vec{k})$, defined as the Fourier transform of the anisotropic stress --- the transverse, traceless projection of the energy-momentum tensor. The source is modeled as an isotropic and homogeneous random variable satisfying
\be
\label{eq:UETC}
    \left\langle \Pi_{ij}(\tau_1, \vec{k}) \Pi_{i j }^*(\tau_2, \vec{k}')  \right  \rangle \! = \! (2\pi)^3 \delta(\vec{k} \! - \!
    \vec{k}') \Pi(\tau_1, \!\tau_2,\! k)\,. ~~~~
\ee
We also assume the transition completes within a fraction of a Hubble time $\beta^{-1} < H_*^{-1}$, where $\beta^{-1}$ is the duration of the PT. This justifies neglecting expansion during the PT, allowing us to drop the Hubble friction term in Eq.~(\ref{eq:fullwaveeq}).\footnote{Interestingly, even for quite long PTs with $\beta/H_* \sim \text{a few}$, incorporating expansion has been shown to have at most an $\mathcal{O}(1)$ effect on the resulting spectrum for this class of PTs~\cite{Yamada:2025long,Yamada:2025short}. Thus, for the purposes of making qualitative forecasts of supercooled PTs, it is justified to estimate the FLRW spectrum from flat-space results.} Finally letting $x \equiv k \tau$, the wave equation during the PT $x_i \leq x \leq x_f$ simplifies to
\begin{equation}
    h_{ij}'' + h_{ij} \simeq \frac{8 \pi G a_*^2}{k^2} \Pi_{ij} \,,
\end{equation}
where now primes denote derivatives with respect to $x$ and $a_* = a(\tau_*)$ is evaluated at the ``time'' of the PT. The solution is
\begin{equation}\label{eq:hijduring}
    h_{ij}(x \leq x_f) = \frac{8 \pi G a_*^2}{k^2} \int_{x_i}^{x} dy \, \sin (x-y) \, \Pi_{ij}(y) \,.
\end{equation}

After the PT completes, the source term is no longer active so the right-hand side of Eq.~(\ref{eq:fullwaveeq}) vanishes, and we can no longer neglect expansion. Assuming radiation domination $\mathcal{H} = 1/\tau$, the wave equation for $x \geq x_f$ is then
\begin{equation}
    h_{ij}'' + \frac{2}{x} h_{ij}' + h_{ij} = 0 \,,
\end{equation}
which has the general solution 
\begin{equation}\label{eq:hijafter}
    h_{ij}( x \geq x_f) = A_{ij} \frac{\sin(x-x_f)}{x} + B_{ij} \frac{\cos(x - x_f)}{x} \,.
\end{equation}
Matching Eqs.~(\ref{eq:hijduring}) and~(\ref{eq:hijafter}) at $x_f$ allows us to identify 
\be
\label{eq:Aij}
    A_{ij} &=& \frac{B_{ij}}{x_f} + \frac{8 \pi G a_*^2}{k^2} \, x_f \int_{x_i}^{x_f} dy \, \cos(x_f - y) \, \Pi_{ij}(y) \,,~~~~ \\
    \label{eq:Bij}
    && B_{ij} = \frac{8 \pi G a_*^2}{k^2} \, x_f \int_{x_i}^{x_f} dy \, \sin(x_f - y) \, \Pi_{ij}(y) \,.
\ee
Given these solutions, we can now define $h^{\rm ini}_{ij}$. Note, however, that because of subtleties involving super-horizon modes, it will be necessary to differentiate between the super-horizon $x_f < 1$ and sub-horizon $x_f > 1$ regimes.\\

\noindent \textbf{Sub-Horizon Regime:}
For sub-horizon modes, we can simply evaluate the initial spectrum at the end of the PT, defining $h_{ij}^{\rm ini} \equiv h_{ij}(x_f) = B_{ij}/x_f$. 
Inserting this solution into Eq.~(\ref{eq:Ph}), the initial power spectrum is\footnote{We have dropped the term $\propto \cos[2(\tau_f - \tau_1 - \tau_2)]$, whose integral vanishes in the infinite time limit adopted by Ref.~\cite{Jinno:2016}.}
\be
\label{eq:Ph_subhor}
\hspace{{-0.05cm}}
    \mathcal{P}_h(k) \simeq\! 16 G^2 a_*^4 k \! \! 
     \int_{\tau_i}^{\tau_f} \! \! \! \! \! d\tau_1  \!\! \int_{\tau_i}^{\tau_f}  \! \! \! \! \!  d\tau_2   \cos [k(\tau_1 \! -\! \tau_2)]  \Pi(\tau_1, \! \tau_2, k) .~~~~~\,
\ee
The double integral here has the same form as Eq.~(23) of Ref.~\cite{Jinno:2016}, which identified the result with the dimensionless function
\be\label{eq:Delta}
\begin{split}
    \Delta\left( \! \frac{k}{a_*\beta} \! \right) & =  \frac{16 G^2 a_*^2 \beta^2 k^3}{3 \kappa^2 H_*^4} \left(  \! \frac{1+\alpha}{\alpha} \! \right)^2 \\
    & \times \int_{t_i}^{t_f}  \!\!dt_1 \int_{t_i}^{t_f} \!\! dt_2  \cos[k(t_1 \! - \! t_2)] \,\Pi(t_1,t_2,k),~~~~~
\end{split}
\ee
where $\kappa$ is the efficiency factor characterizing the fraction of vacuum energy converted to bulk kinetic energy. Note that $\Delta$ depends only on the ratio $k/(a_*\beta)$ and can be approximated as \cite{Jinno:2016}
\begin{equation}\label{eq:delta-approx}
    \Delta(\tilde{k}) \simeq \Delta_{p}  \bigl[  {c_l \tilde{k}^{-3} + (1 - c_l - c_h) \tilde{k}^{-1} + c_h \tilde{k} } \bigr]^{-1} \,,
\end{equation}
where $\tilde{k} \equiv k/k_p$, $k_{p} \simeq 1.24 \, (a_* \beta)$ is the peak frequency of the PT, $\Delta_{p} \simeq 0.043$  is the peak amplitude, $c_l \simeq 0.064$, and $c_h \simeq 0.48$. This approximation is accurate to within  $8 \%$ of the full function, derived in Sec.~II of the Supplemental Materials.

In terms of $\Delta$, the tensor power spectrum in the sub-horizon regime becomes
\be\label{eq:Pini}
   \!\! \mathcal{P}_h^{\rm sub}(k) = 3 \kappa^2 \! \left( \! \frac{\alpha}{1+\alpha} \! \right)^2 \!\! \left( \frac{a_* H_*}{k} \right)^{\! 2} \!\!\left( \frac{H_*}{\beta} \right)^{\! 2} 
    \!
    \Delta \! \left( \! \frac{k}{a_*\beta} \! \right)\! .~~~
\ee
From Eq.~(\ref{eq:delta-approx}), we see $\Delta(y \gg 1) \simeq 0.11/y$ in the sub-horizon regime. Thus
\begin{equation}\label{eq:subPh}
    \mathcal{P}_h^{\rm sub}(k) \approx 0.33 \, \kappa^2 \left( \frac{\alpha}{1 + \alpha} \right)^2 \left( \frac{H_*}{\beta} \right) \left( k \tau_* \right)^{-3} \,,
\end{equation}
where in the last step we have used $1/(a_* H_*) = \tau_*$.\\

\noindent \textbf{Super-Horizon Regime:}
Activating a super-horizon mode is analogous to exciting an overdamped oscillator~\cite{Hook:2020}; it takes a small but finite amount of time $\epsilon \ll 1$ for the mode to grow to its maximum amplitude, which then remains constant until horizon re-entry at $x = 1$. Since the mode needs to have grown to its maximum amplitude before we can safely apply the transfer function, for super-horizon modes we should define $h_{ij}^{\rm ini} \equiv h_{ij}(x_f + \epsilon)$, with $h_{ij}$ in Eq.~(\ref{eq:hijafter}).

\begin{figure}[t!]
\hspace{-0.4cm}
\includegraphics[width=0.48\textwidth]{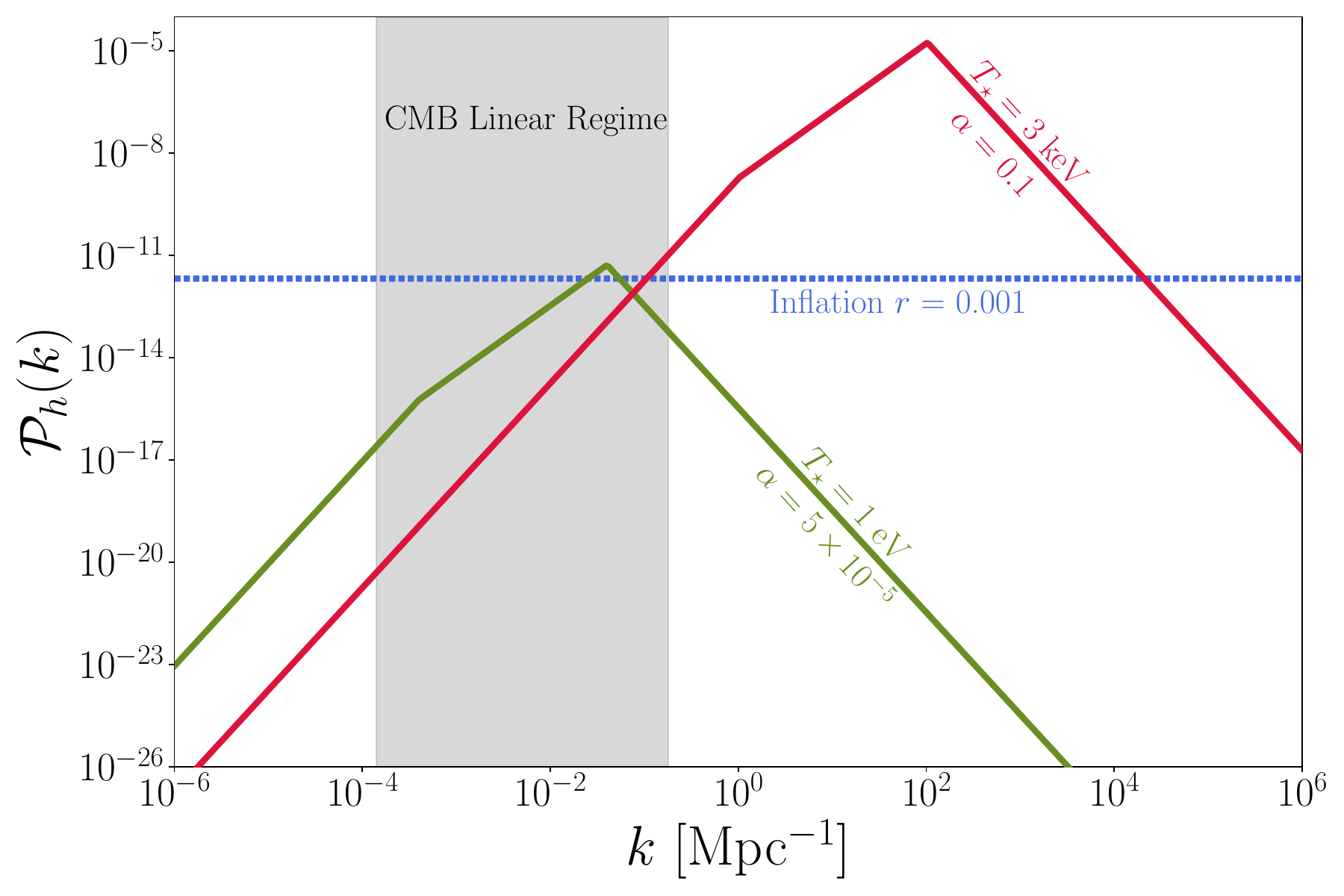}
\caption{Tensor power spectra from Eq.~(\ref{eq:Phfull}) for two sample cosmological PTs with $\kappa=1$, $\beta/H_*=3$, and $\epsilon_b = 10^{-2}$. Note that these parameters are compatible with other limits on cosmological PTs, although such limits are model-dependent and vary according to the post-transition equation of state (see Sec.~IV of the Supplemental Materials for further discussion). We compare with an inflationary prediction for $r = 10^{-3}$ (dashed blue), which is within the projected sensitivity of CMB-S4~\cite{CMB-S4:2016ple}. The gray shaded region represents the linear regime probed by primary CMB anisotropy measurements ($1<\ell<2500$).}
\label{fig:Ph}
\end{figure}
\begin{figure*}[t!]
\centering
\hspace{-0.5cm}\includegraphics[width=0.5\textwidth]{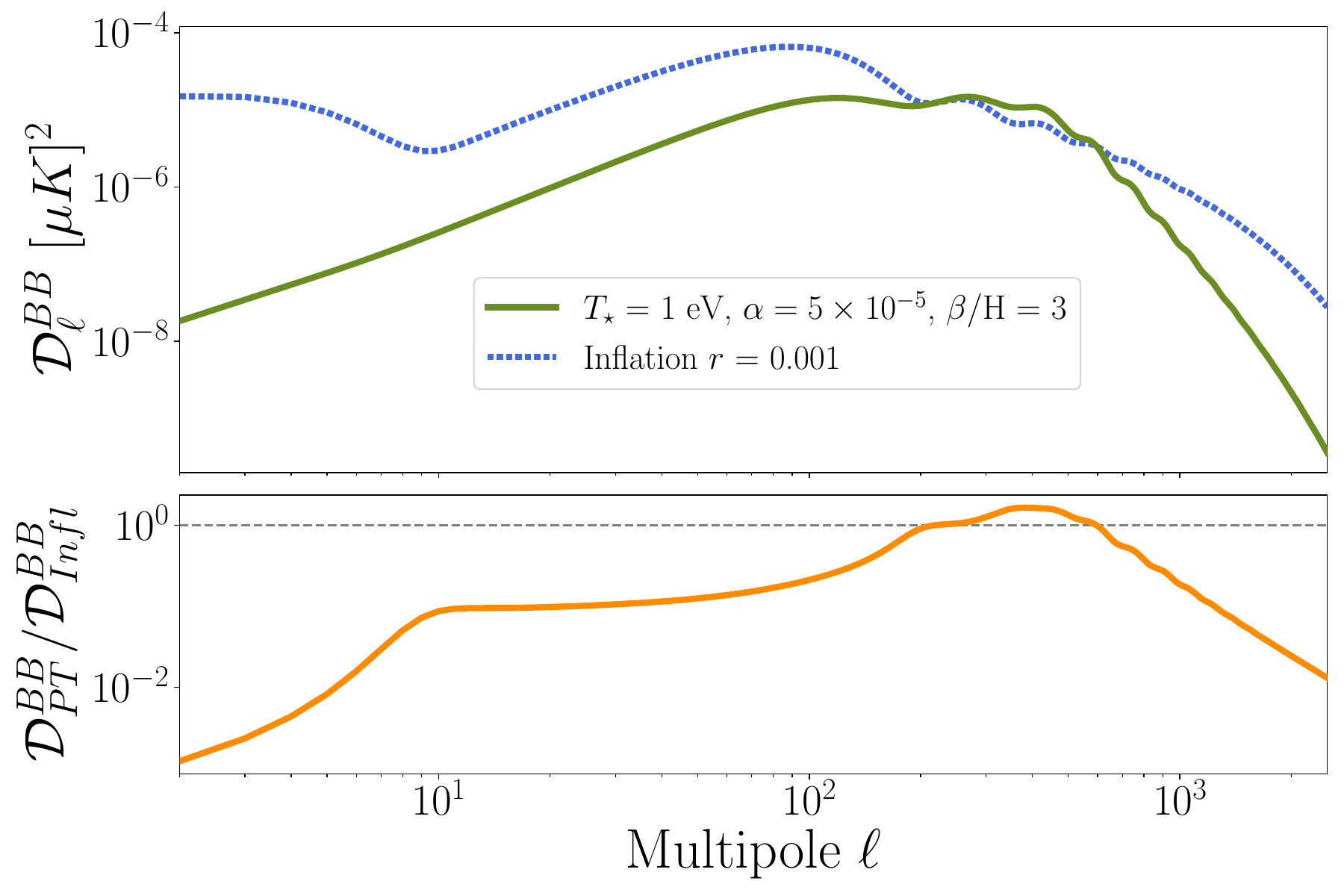}~~~~~~~
\hspace{-0.5cm}\includegraphics[width=0.5\textwidth]{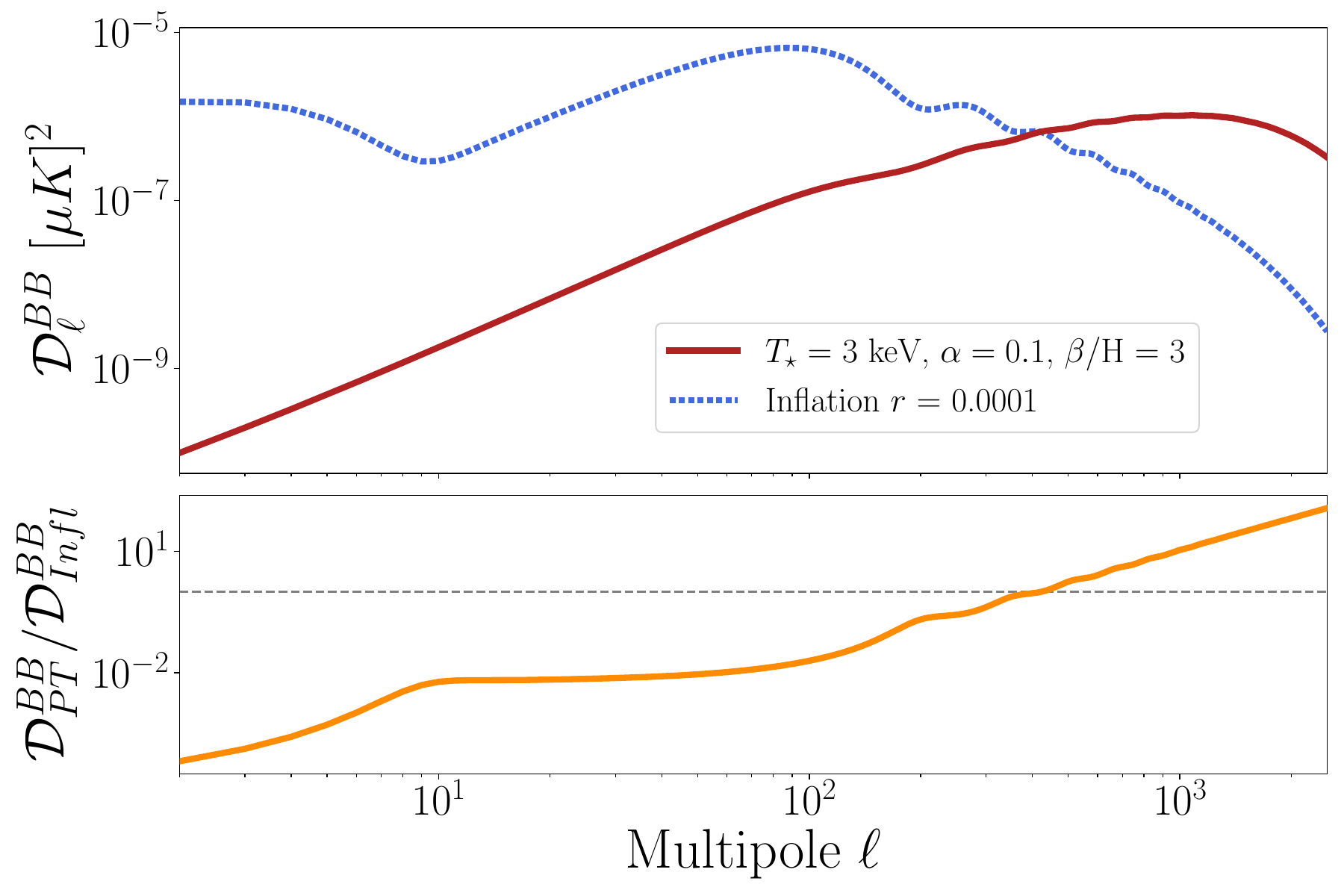}~
\caption{\textbf{Top Left}: $B$-mode polarization spectra for the green phase transition benchmark of Fig.~\ref{fig:Ph} (solid green) and a minimal inflationary model with $r=0.001$ (dashed blue). Note that lensing has been removed.  
{\bf Bottom Left}:
Ratio of $B$-mode signals from these sources. 
\textbf{Right}: Same as the left panel, but for the red benchmark of Fig.~\ref{fig:Ph}. We compare with an inflationary scenario with $r = 10^{-4}$, chosen to match the most optimistic projections of CMB-S4~\cite{CMB-S4:2016ple}.
}
\label{fig:DlBB}
\end{figure*}

Because $\beta/H_* > 1$ and $x_f - x_i \simeq x_f (H_*/\beta)$, Eqs. \eqref{eq:Aij} and \eqref{eq:Bij} scale as
\be
\label{eq:scaling}
A_{ij} \propto  x_f^2 (H_*/\beta) \,,~~ B_{ij} \propto x_f^3(H_*/\beta)^2 \,.~~
\ee
Since $x_f \ll 1$ in the super-horizon limit, we generically have $A_{ij} \gg B_{ij}$. Furthermore, just after the super-horizon mode has acquired its maximum value, 
the terms in Eq.~(\ref{eq:hijafter}) become 
\be
 A_{ij}\frac{\sin(x-x_f)}{x} \to A_{ij} \,,~~ B_{ij}\frac{\cos(x-x_f)}{x} \to  
 \frac{B_{ij}}{x}  \,,~~~
\ee
so only the first term survives in this limit and $h^{\rm ini}_{ij} = A_{ij}$. Inserting this solution into Eq.~(\ref{eq:Ph}), the initial power spectrum is
\begin{equation}
    \mathcal{P}^{\rm super}_h(k) = 3 \kappa^2 \left( \frac{\alpha}{1 + \alpha} \right)^2 \left( \frac{H_*}{\beta} \right)^2 \Delta \left( \frac{k}{a_* \beta} \right) \,,
\end{equation}
where we have again used the definition of $\Delta$
from \Eq{eq:Delta}. Deep in the super-horizon regime, Eq.~(\ref{eq:delta-approx}) can be approximated as $\Delta(y \ll 1) \simeq 0.352 y^3$, and so
\begin{equation}\label{eq:superPh}
    \mathcal{P}^{\rm super}_h(k) \simeq 1.1 \, \kappa^2 \left( \frac{\alpha}{1 + \alpha} \right)^2 \left( \frac{H_*}{\beta} \right)^5 \left( k \tau_* \right)^3 \,.
\end{equation}
Eq.~(\ref{eq:superPh}) exhibits the expected $k^3$ scaling characteristic of causality-limited modes, whose wavelengths are much larger than the spatial correlations of the source~\cite{Caprini:2009fx,Cai:2019,Hook:2020}. For such modes, the power spectrum is simply white noise and gets strongly suppressed since $(k \tau_*) \ll 1$ outside the horizon. See Sec.~III of the Supplemental Materials for more detail.\\

\noindent \textbf{Full Spectrum:} The power spectrum peaks at the maximal bubble size, $k_p \simeq 1.24 (\beta/H_*) \tau_*^{-1}$~\cite{Jinno:2016}. For $k \geq k_p$, $\mathcal{P}_h$ is described by the sub-horizon spectrum of Eq.~(\ref{eq:subPh}); for $k \ll k_p$, $\mathcal{P}_h$ is described by the super-horizon spectrum of Eq.~(\ref{eq:superPh}). The intermediate regime $k \lesssim k_p$ corresponds to modes which are sub-horizon but super-bubble, and exhibits greater theoretical uncertainty. We parameterize our ignorance about this regime by defining a ``breaking scale''
\begin{equation}
    k_b \equiv \epsilon_b k_p \,, \,\,\,\,\,\, \epsilon_b \ll 1 \,,
\end{equation}
where the $k^3$ super-horizon scaling breaks down. For modes in this range $k_b \leq k \leq k_p$, we model the spectrum as a power law
\be
    \mathcal{P}_h^{\rm int}(k) = A k^m \,, 
\ee
whose slope and amplitude satisfy 
\be
 m = \frac{\log \left( \mathcal{P}_h^p/\mathcal{P}_h^b \right)}{\log \left( k_p / k_b \right)}~~,~~ A = \frac{\mathcal{P}_h^b}{k_b^m} \, ,
\ee
to match the boundary values $\mathcal{P}_h^p \equiv \mathcal{P}_h(k_p)$ and $\mathcal{P}_h^b \equiv \mathcal{P}_h(k_b)$. The full tensor power spectrum is then
\begin{equation}\label{eq:Phfull}
    \mathcal{P}_h(k) = \begin{cases} \mathcal{P}^{\rm super}_h(k) & k \leq k_b \,, \\ \mathcal{P}_h^{\rm int}(k) & k_b \leq k \leq k_p \,, \\ \mathcal{P}^{\rm sub}_h(k) & k \geq k_p \,, \end{cases}
\end{equation}
which we plot in Fig.~\ref{fig:Ph} for two sample benchmarks. We comment that the benchmark parameter values feature small values of $\alpha$, which are atypical for strongly-supercooled PTs. These small values are chosen to maintain consistency with various bounds on late-time PTs (see Sec.~IV of the Supplemental Materials). One would typically expect a larger contribution from sound waves in this parameter regime, which we do not calculate. In this sense, our estimate of the signal is conservative.

\section{Numerical Results}
\label{sec:numerical}

Fig.~\ref{fig:DlBB} shows the angular $B$-mode power spectrum 
\begin{equation} 
\label{eq:DlBB}
    \mathcal{D}_\ell^{BB} \equiv  \frac{\ell (\ell+1)}{2\pi} T_0^2 C_\ell^{BB} \,,
\end{equation} 
for sample PT and inflationary scenarios. We exclude the predicted lensing signal from all predictions to highlight the intrinsic shape differences between these sources. Note that the lensing signal is a foreground effect that impacts both models in a similar way \cite{Seljak:1995ve,Bernardeau:1996aa,Zaldarriaga:1998ar,Hu:2001kj,Hu:2001fb,Hu:2002vu,Seljak:2003pn,Lewis:2006fu,Hanson:2009kr,Dodelson:2010qu,Hotinli:2021umk,Trendafilova:2023xtq}.

The curves in these figures are computed using the Boltzmann solver \texttt{CLASS} \cite{Blas:2011}. For the PT signal, we utilize the \texttt{external-pk} module with the custom primordial tensor power spectrum from \Eq{eq:Phfull}. \texttt{CLASS} computes the transfer functions from the Einstein-Boltzmann equations and the polarization source functions. These outputs, combined with the primordial tensor power spectrum, are then used to evaluate the line-of-sight and $k$-space integrals to obtain the $\mathcal{D}_{\ell}^{BB}$ spectrum. 

We remark that for both benchmarks, the PT has a distinct spectral shape as compared with the inflationary predictions. Notably, the PT spectra peak at higher multipoles than inflationary spectra, which peak around $\ell \sim 100$. This behavior can be understood from the definition of $C_\ell^{BB}$ in Eq.~(\ref{eq:ClBB}) and the shape of the power spectrum for each scenario. Note that the domain of support for the function $\mathcal{F}_\ell(k)$ roughly corresponds to the gray region in Fig.~\ref{fig:Ph}. $\mathcal{F}_\ell(k)$ peaks around $k \sim 0.01 \, \text{Mpc}^{-1}$, decreases slowly on larger scales, and decreases very quickly on smaller scales. 

 The inflationary power spectrum is nearly scale invariant $\mathcal{P}_h \sim k^0$, so the dominant contribution to the integral in Eq.~(\ref{eq:ClBB}) comes from modes in the vicinity of the peak in $\mathcal{F}_\ell(k)$. By using the approximation $\ell \sim k \tau_0$, one can show that $k \sim 0.01\, \text{Mpc}^{-1}$ roughly corresponds to a maximal $B$-mode signal at $\ell \sim 100$, consistent with  Fig.~\ref{fig:DlBB}. By contrast, the PT signal exhibits maximal power on smaller scales $k > 0.01 \, \text{Mpc}^{-1}$ and becomes sharply suppressed on large scales since $\mathcal{P}_h(k) \propto k^{3}$ in the infrared tail. Thus, the dominant contribution to the integral in Eq.~(\ref{eq:ClBB}) comes from these larger $k$ values, despite the suppression in $\mathcal{F}_\ell(k)$ that occurs here. 

\begin{figure}[t!]
\hspace{-0.4cm}
\includegraphics[width=0.48\textwidth]{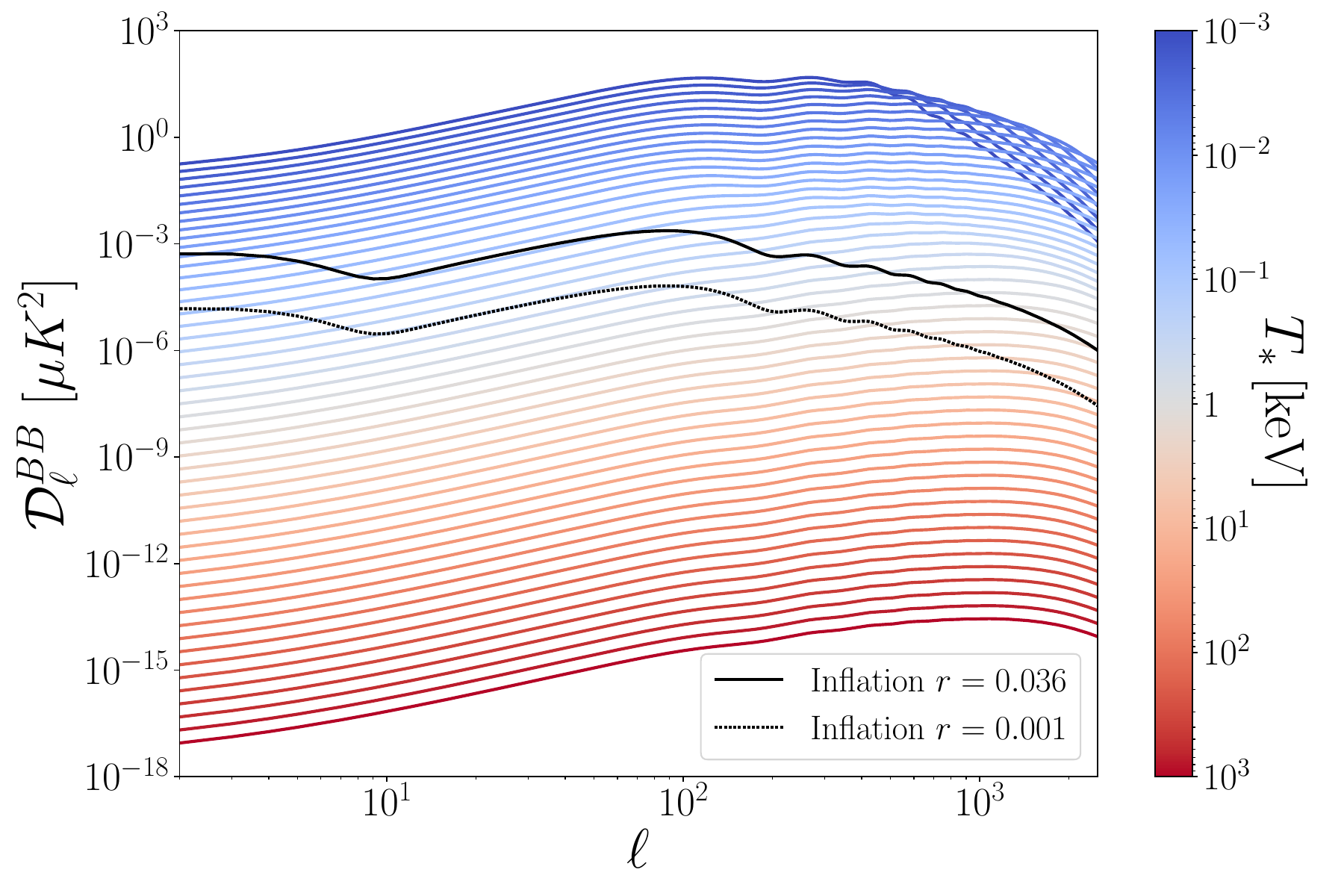}
\caption{Color map of $B$-mode power spectra for various $T_*$ fixing $\kappa = 1$, $\beta/H_* = 3$, and $\alpha = 0.1$. Solid black curve: Inflationary prediction for $r = 0.036$, which saturates current BICEP limits \cite{BICEP:2021xfz}. Dotted black curve: Inflationary prediction for $r = 10^{-3}$, 
corresponding to CMB-S4 sensitivity projections \cite{CMB-S4:2016ple}.} 
\label{fig:color} 
\end{figure}

For the $T_* = 1\, \text{eV}$ benchmark, the integral in Eq.~(\ref{eq:ClBB}) receives contributions from all three regions in \Eq{eq:Phfull}, giving rise to both low- and high-$\ell$ suppression relative to the inflationary signal. Meanwhile for the $T_* = 3\, \text{keV}$ benchmark, the integral samples only from the super-horizon tail. Because $\mathcal{P}_h \propto k^3$ here, $\mathcal{D}_\ell^{BB}$ exhibits low-$\ell$ suppression and high-$\ell$ enhancement relative to the inflationary signal. The transition in the shape of $\mathcal{D}_\ell^{BB}$ as one begins sampling from all three regions can also be seen in the low-$T_*$ contours of Fig.~\ref{fig:color}.

\section{Complementary GW Signal}\label{sec:OmegaGW}

The tensor perturbations sourcing CMB $B$-mode polarization also contribute to the stochastic GW background at late times. The relative energy density in GWs is quantified by the spectral density parameter
$\Omega_{\rm GW} \equiv \rho^{-1}_{\rm tot} (d\rho_{\rm GW}/d\log k)$.
At  $\tau = \tau_*$, the energy density in GWs is
\begin{equation}\label{eq.omegaGWst}
    \rho_{\rm GW}^* = \frac{1}{8 \pi G a_*^2} \int d \ln k \, \mathcal{P}_{h'}(\tau,k) \,,
\end{equation}
where $\mathcal{P}_{h'}(k)$ is the dimensionless power spectrum of the derivative of the tensor perturbation with respect to $\tau$. In calculating $\mathcal{P}_{h'}(k)$, there is no longer any subtlety for the super-horizon modes, since the initial condition for the derivative of the field amplitude is non-vanishing. Thus, we calculate
$(h_{ij}')^{\rm ini} \equiv h_{ij}' \big|_{\tau_f}$ from Eq.~(\ref{eq:hijduring}) and compute the power spectrum at $\tau = \tau_f$ to obtain 
\begin{equation}\label{eq:OmegaGWstar}
    \Omega_{\rm GW}^* = \kappa^2 \left( \frac{\alpha}{1 + \alpha} \right)^2 \left( \frac{H_*}{\beta} \right)^2 \Delta \left( \frac{2 \pi f_*}{\beta} \right) \,,
\end{equation}
where $f = k/(2\pi a)$ is the physical frequency.
To extract the present-day GW signal, we redshift the frequency and the energy density to obtain 
\begin{equation}\label{eq.Omegagw}
\begin{split}
    \Omega_{\rm GW} h^2 \!=\! 8 \!\times \! 10^{-5} \frac{g_\star}{g_{\star,s}^{4/3}} \!\left( \! \frac{\kappa \alpha}{1 + \alpha} \! \right)^2 \!\! \left( \! \frac{H_*}{\beta} \! \right)^2 \!\! \Delta \! \left( \! \frac{2\pi f_0}{a_* \beta} \! \right) \,,
\end{split}
\end{equation}
where $g_\star = g_\star(T_*)$ and $g_{\star,s} = g_{\star,s}(T_*)$ are understood to be evaluated at $T_*$ and in general depend on the number of relativistic degrees of freedom in the dark sector.

\begin{figure}[t!]
\centering
\hspace{-0.4cm}
\includegraphics[width=0.48\textwidth]{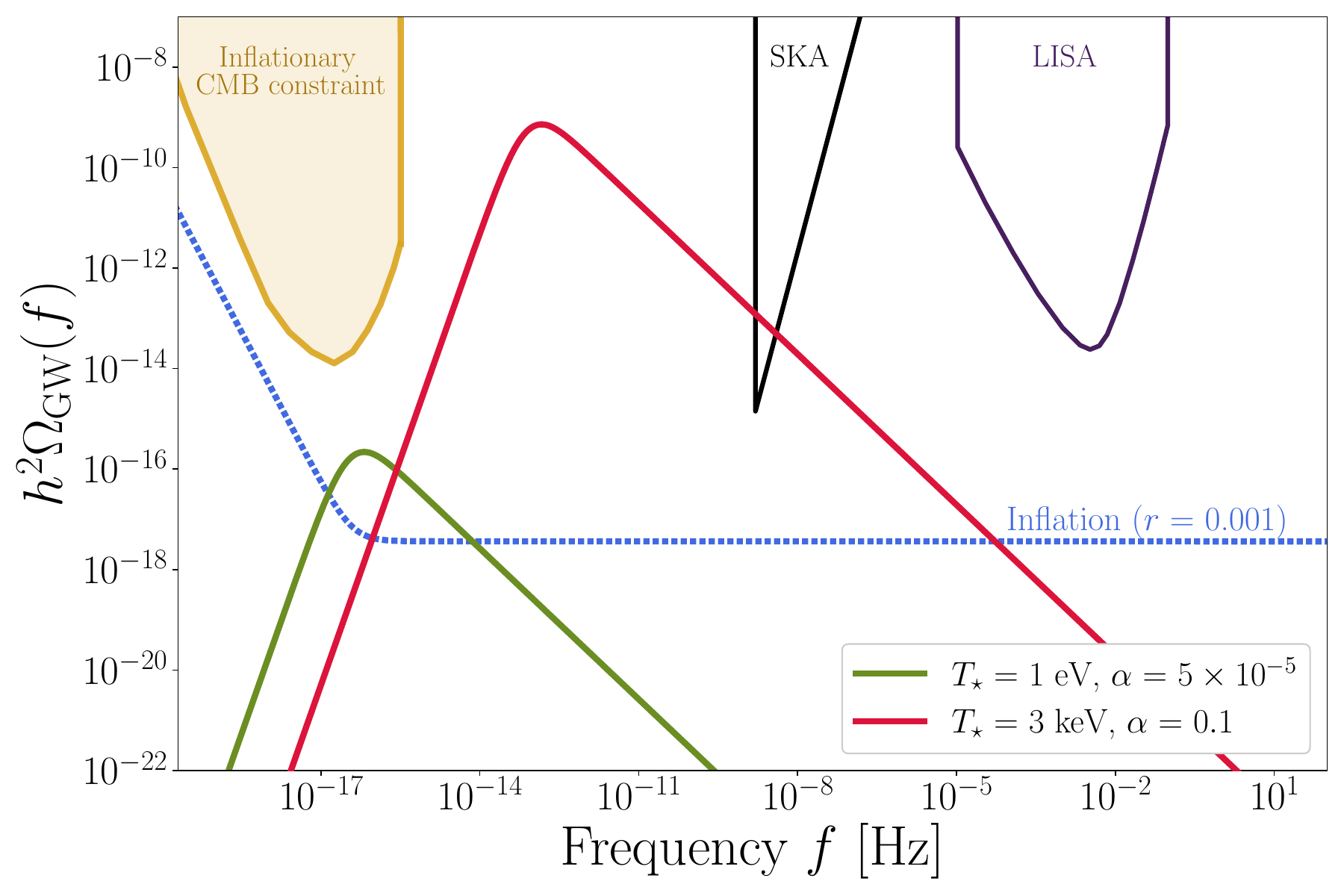}
\caption{Representative GW spectra from phase transitions (solid red/green) and inflation (dotted blue). Parameter values match the benchmarks of Fig.~\ref{fig:Ph}. We assume 1 massless dark photon as the relativistic particle content in the dark sector for concreteness. We show also the CMB constraint on inflation from Ref.~\cite{Lasky:2015lej} (yellow shaded) and projected sensitivities for LISA (purple) \cite{LISA:2017pwj} and SKA (black) \cite{Janssen:2014dka}.
}
\label{fig:GW}
\end{figure}

In Fig.~\ref{fig:GW}, we show sample $\Omega_{\rm GW}$ predictions for the PT and inflationary scenarios, presented alongside the CMB constraints from Ref.~\cite{Lasky:2015lej}. Note that the precise location of the CMB limit from $B$-modes is model dependent, as translating sensitivity from ${\cal D}_\ell^{BB}$ to $\Omega_{\rm GW}$ depends on the power spectrum of the source. Notice also that the PT and inflationary lines cross around $f \sim 10^{-17} \, \text{Hz}$, corresponding to $k \sim 0.01 \, \text{Mpc}^{-1}$ --- exactly as one would expect, given these benchmarks lead to similar maximal amplitudes in the $B$-mode polarization spectrum.

\section{ Discussion}\label{sec:discussion}

In this {\it Letter}, we have found that the tensor perturbations produced during a first-order cosmological phase transition can in principle generate CMB $B$-mode polarization on observable scales. Although phase transitions occur due to causal processes on sub-horizon scales, they nevertheless exhibit a white noise power spectrum on large scales.\footnote{This form for the power spectrum follows from causality and implies that fluctuations at sufficiently separated spatial points are uncorrelated. See Sec.~III of the Supplemental Materials.} Power is suppressed on the scales relevant for the CMB, but it is not necessarily negligible. For sufficiently strong late-time transitions, the amplitude can be comparable to inflationary predictions which can be probed with CMB-S4~\cite{CMB-S4:2016ple}. Thus, a discovery of $B$-modes beyond the known lensing effect is no longer a definitive  ``smoking gun'' for inflation, though it is still clear evidence of new physics. 

Fortunately, the spectral shape of the PT signal is distinct from inflationary predictions. While the latter peaks around $\ell \sim 100$, the PT generically predicts peak power at smaller scales (higher $\ell$). Thus, the origin of a potential signal can ultimately be distinguished with sufficiently precise $B$-mode measurements across multiple angular scales. Furthermore, the cosmological PTs we have considered may also predict stochastic GW backgrounds within SKA sensitivity \cite{Janssen:2014dka}, CMB spectral distortions observable with PIXIE \cite{Kogut:2024vbi} and SPECTER \cite{Sabyr:2024lgg}, and  $\Delta N_{\rm eff}$ within the reach of CMB-S4 targets \cite{CMB-S4:2016ple}.

Our work has focused on tensor modes from the bubble collision stage of a first-order PT occurring in a thermally-decoupled dark sector as a proof-of-principle. There are several ways in which one could improve upon the estimates presented here. First, while Refs.~\cite{Yamada:2025long,Yamada:2025short} show that the flat-space calculation suffices for qualitative predictions of supercooled PTs, for the longest PTs the error incurred by neglecting expansion can become an $\mathcal{O}(1)$ factor, and so should be accounted for in making precise predictions for concrete models. Additionally, the parameter values explored here, which were chosen to maintain consistency with bounds on late-time PTs, one should expect the dominant contribution to come from sound waves. Since we have not included this contribution, these estimates are conservative. Future work should endeavor to include tensor perturbations from the sound wave and turbulence stages as well, as fully accounting for these effects could allow for much louder signals within viable parameter space. 


Finally, there are many other mechanisms which can source tensor perturbations in the post-inflationary universe, whose near-term observational prospects may be more optimistic. These include cosmic string networks, domain walls, scalar-induced gravitational waves, and more. We leave the detailed study of such sources to future work. 


While our result underscores the challenge of understanding inflation from a $B$-mode discovery, it also highlights the exciting opportunity such a discovery affords for uncovering a broader landscape of new physics. This motivates not only the exploration of novel mechanisms for generating tensor modes, but also the development of new experimental strategies aimed at detecting higher-$\ell$ B-modes.

\section*{Acknowledgements}
We would like to thank Adam Anderson, Francis-Yan Cyr-Racine, Scott Dodelson, Josh Foster, Peter Graham, Wayne Hu, Marc Kamionkowski, Soubhik Kumar, Gustavo Marques-Tavares, Robert McGehee, Jeff McMahon, Julian Mu\~noz, Albert Stebbins, Matthew Young, and Jessica Zebrowski for helpful conversations. AI is supported by NSF Grant PHY-2310429, Simons Investigator Award No.~824870, DOE HEP QuantISED award \#100495, the Gordon and Betty Moore Foundation Grant GBMF7946, and the U.S.~Department of Energy (DOE), Office of Science, National Quantum Information Science Research Centers, Superconducting Quantum Materials and Systems Center (SQMS) under contract No.~DEAC02-07CH11359. Fermilab is operated by the Fermi Research Alliance, LLC under Contract DE-AC02-07CH11359 with the U.S. Department of Energy. This material is based partly on support from the Kavli Institute for Cosmological Physics at the University of Chicago through an endowment from the Kavli Foundation and its founder Fred Kavli. This material is based upon work supported by the U.S. Department of Energy, Office of Science, Office of Workforce Development for Teachers and Scientists, Office of Science Graduate Student Research (SCGSR) program. The SCGSR program is administered by the Oak Ridge Institute for Science and Education for the DOE under contract number DE‐SC0014664. YT is supported by the NSF Grant PHY-2112540 and PHY-2412701. YT would like to thank the Tom
and Carolyn Marquez Chair Fund for its generous support. YT would also like to thank the
Aspen Center for Physics (supported by NSF grant PHY-2210452).

\appendix

\section{$B$-mode Polarization Signal}\label{sec:ClBB}

When photons scatter with free electrons, quadrupole anisotropies in the temperature distribution are transformed into polarization of the scattered photons. The majority of CMB polarization is generated during recombination, since afterwards the number density of free electrons drops sharply and Thomson scattering ceases to be efficient. Here, we make the simplifying assumption that all polarization is generated in this last scattering event. This approximation is relaxed in the results of the main text, where we compute the angular $B$-mode spectrum exactly using the Boltzmann solver {\tt CLASS} \cite{Lesgourgues:2011,Blas:2011}. The purpose of these formulas is simply to provide intuition as well as a semi-analytic check of our results.

Consider initially unpolarized photons which arrive along direction $\hat{n}'$ to the point $\vec{x}$, where they last scatter at conformal time $\tau$. Letting $\hat{n}$ be the direction of observation and $\tau_0$ be the conformal time today, we can write $\vec{x} = (\tau_0 - \tau)\hat{n}$. A tensor perturbation $h_{ij}(\tau,\vec{k})$ gives the following contribution to the temperature anisotropy $\Theta \equiv \Delta T/T$, which can be written \cite{Rubakov:2017}
\begin{equation}
\begin{split}
    \Theta (&\tau, \vec{k}; \hat{n}, \hat{n}') = \frac{1}{2} \int_0^\tau d\tau_1 
    \, V(\tau,\tau_1) 
    e^{i\vec{k} \cdot \hat{n} (\tau_0 - \tau)} \\
    &\times 
   \int_{\tau_1}^{\tau} d\tau_2
    e^{- i \vec{k} \cdot \hat{n}' (\tau - \tau_2)}  \!
 \sum_{\lambda= +, \times}  n_i' \epsilon_{ij}^{\lambda} n_j' \, 
 \partial_{\tau_2} h_{\lambda}(\tau_2, \vec{k})
\end{split}
\end{equation}
where $\epsilon_{ij}^{\lambda}$ is the polarization tensor for gravitational waves, the sum runs over graviton polarizations, and $V(\tau_1, \tau_2)$ is the visibility function, which satisfies
\begin{equation}
    V(\tau_1, \tau_2) = e^{-\kappa(\tau_1,\tau_2)} \frac{d\kappa(\tau_2)}{d\tau_2} \,,
\end{equation}
and is defined in terms of the optical depth
\begin{equation}
    \kappa(\tau_1, \tau_2) = \int_{\tau_2}^{\tau_1} d\tau \, a(\tau) \, \sigma_T \, n_e(\tau) \,
\end{equation}
where $n_{\rm e}$ is the electron number density and $\sigma_{\rm T}$ is the Thomson cross section.
This temperature anisotropy gives rise to an anistropy in the intensity of incoming radiation, and so $\Theta$ enters into the CMB polarization tensor as \cite{Rubakov:2017,Kosowsky:1994}
\be
      P_{ab}(\vec{k}; \hat{n})  = \frac{3}{4\pi} \!\int d^2n' 
      \varepsilon_{ab} \!
    \int_0^{\tau_0} \!\! d\tau  V(\tau_0,\tau) \Theta (\tau, \vec{k}; \hat{n}\cdot\hat{n}'),~~~~
\ee
and we have defined the tensor
\be
\varepsilon_{ab} = \frac{1 - (\hat{n} \cdot \hat{n}')^2}{2} g_{ab} - \hat{n}' \mathbf{e}_a \cdot \hat{n}' \mathbf{e}_b~,
\ee
where $\mathbf{e}_a$ is the set of basis vectors on the celestial sphere and $g_{ab}$ is the background metric. Notice that because $\hat{n}'$ enters in a bilinear combination, polarization requires a quadrupole component to the anisotropy. 

It is convenient to decompose the polarization tensor into $E$- and $B$-modes, which can each be expanded in spherical harmonics on the celestial sphere. Doing so, one can identify the coefficient
\begin{equation}
    a_{\ell m}^B(\vec{k}) = - \int d^2 n \, \big[Y_{\ell m}^{(B)}\big]_{ab}^* (n) \, P^{ab}(\vec{k}; \hat{n}) \,,
\end{equation}
where $Y_{\ell m}^{(B)}$ are the $B$-mode tensor harmonics on the sphere.\footnote{The $B$-mode tensor harmonics are related to the ordinary spherical harmonics as
\begin{equation}
    \big(Y_{\ell m}^{(B)}\big)_{ab}(n) = \sqrt{\frac{(\ell - 2)!}{2 (\ell + 2)!}}\left( \epsilon_b^c \nabla_a \nabla_c + \epsilon_a^c \nabla_c \nabla_b \right) Y_{\ell m}(n) \,.
\end{equation}
}
The angular spectrum of $B$-mode polarization is defined in terms of these coefficients as
\begin{equation}
\label{eq:CellBB}
    C_\ell^{BB} = \frac{1}{2\ell + 1} \sum_{m = \pm 2} \int \frac{d^3 k}{(2\pi)^3} \, \expval{a_{\ell m}^B(\vec{k})\, a_{\ell m}^{B}(\vec{k})^*} \,,
\end{equation}
where the angle brackets denote ensemble average. Working in a frame with the azimuthal axis oriented along $\vec{k}$ to perform the integrals, one can show that
\begin{equation}\label{eq:ClBB}
    C_\ell^{BB} = 36 \pi \int_0^\infty \frac{dk}{k} \mathcal{P}_h(k) \mathcal{F}_\ell(k)^2 \,,
\end{equation}
where $\mathcal{P}_h(k)$ is the dimensionless tensor power spectrum evaluated at the initial time and 
\be\label{eq:Fl}
    \mathcal{F}_\ell (k) &=& \int_0^{\tau_0} d\tau \, V(\tau_0,\tau) \, {\cal S}_\ell(k, \tau_0, \tau)      \int_0^\tau d\tau_1 \, V(\tau,\tau_1)  \nonumber
    \\ 
    && ~~~\times
  \int_{\tau_1}^{\tau} d\tau_2  \, \frac{j_2[k(\tau - \tau_2)]}{k^2 (\tau - \tau_2)^2} 
  \brac{\partial \mathcal{T}(\tau_2,k)}{\partial \tau_2}
  ,~~~~~ 
\ee
where we have defined
\be
\label{eq:Sfunc}
{\cal S}_\ell(k, \tau_0, \!\tau) \equiv  \frac{\ell \!+\! 2}{2 \ell \! +\! 1} j_{\ell - 1} \!\left[ k(\tau_0 \!-\! \tau) \right] \!- \!\frac{\ell \!-\! 1}{2\ell \!+\! 1} j_{\ell + 1}\! \left[ k (\tau_0 \!-\! \tau) \right] \! .~~~~~~ 
\ee
Note that in deriving this result, we have decomposed the tensor perturbation into an initial perturbation amplitude $h^{\rm ini}_\lambda(k)$ and the transfer function $\mathcal{T}(\tau,\vec{k})$, where 
\begin{equation}
    h_\lambda(\tau, \vec{k}) = h_\lambda^{\rm ini}(\vec{k}) \mathcal{T} (\tau, \vec{k}) \,.
\end{equation}
This decomposition is useful because it separates the effect of statistical correlations between the initial amplitudes from the deterministic effect of the modes' subsequent evolution, as captured by the transfer function. The statistical properties of the initial perturbations are encoded in the (dimensionful) power spectrum $P_h(k)$,
\begin{equation}\label{eq:app-Ph}
    \expval{h^{\rm ini}_\lambda(\vec{k}) \, h^{\rm ini}_{\lambda'}(\vec{k}')^*} = \frac{\delta_{\lambda \lambda'}}{2} P_h(k) (2\pi)^3 \delta^{(3)}(\vec{k} - \vec{k}') \,.
\end{equation}
We also introduce the dimensionless power spectrum appearing in Eq.~(\ref{eq:ClBB}),
\be
\label{eq:mathcalPh}
    \mathcal{P}_h(k) = \frac{k^3}{2\pi^2} P_h(k) \,.
\ee

In the presence of a nonzero source $\Pi_{ij}(\tau, \vec{k})$, a Fourier mode of the metric perturbation $h_{ij}(\tau,\vec{k})$ evolves according to the wave equation
\begin{equation}\label{eq:fullwaveeq}
    h_{ij}'' + 2 \mathcal{H} h_{ij}' + k^2 h_{ij} = 8 \pi G a^2 \Pi_{ij} \,,
\end{equation}
where $\mathcal{H} = a'/a$ is the conformal Hubble rate, primes denote derivatives with respect to conformal time, and the source term is the Fourier transformed  anisotropic stress. When the source is inactive, the transfer function satisfies
\begin{equation}
    \mathcal{T}'' + 2 \mathcal{H} \mathcal{T}' + k^2 \mathcal{T} = 0 \,.
\end{equation}
Following Ref.~\cite{Kite:2021}, we ignore the late-time contribution from dark energy, and consider only effects from the transition from radiation to matter domination. In this 2-component universe, the Friedman equations can be solved analytically, and one can derive the following solutions for the transfer function in the radiation-dominated and matter-dominated regimes
\be
\label{eq:TRD}
    \mathcal{T}_{\rm RD}(\tau,k) &=& A_k \, j_0(k \tau) - B_k \, y_0(k \tau) \\ 
    \label{eq:TMD}
     \mathcal{T}_{\rm MD}(\tau,k) &=& \frac{3}{k \tau} \big[ C_k \, j_1(k \tau) - D_k \, y_1(k \tau) \big] \,.
\ee
From the initial conditions and the matching conditions at the characteristic timescale $\tilde{\tau} = 4 \sqrt{\Omega_r}/H_0 \Omega_m$, the constants in \Eq{eq:TRD} are $A_k = 1 ,  B_k = 0$,  and those
 in \Eq{eq:TMD} satisfy
\begin{equation}\label{eq:coeff}
\begin{split}
    & C_k  = \frac{1}{2} - \frac{\cos(2 k \tilde{\tau})}{6} + \frac{\sin(2 k \tilde{\tau})}{3 k \tilde{\tau}} \,, \\
    & D_k  = - \frac{1}{3 k \tilde{\tau}} + \frac{k \tilde{\tau}}{3} + \frac{\cos(2 k \tilde{\tau})}{3 k \tilde{\tau}} + \frac{\sin(2 k \tilde{\tau})}{6} \,.
\end{split}
\end{equation}

Given a form for the initial tensor power spectrum, Eq.~(\ref{eq:ClBB}) in conjunction with Eqs.~(\ref{eq:Fl}), (\ref{eq:TRD}), (\ref{eq:TMD}), and (\ref{eq:coeff}) can be used to compute the $B$-mode signal. These expressions contain many nested integrals with oscillatory integrands, however, which can lead to numerical instabilities. Approximating the visibility function as a Gaussian of width $\Delta \tau_r \simeq 0.04 \tau_r$ about its maximum \cite{Rubakov:2017}, the conformal time integrals can be simplified considerably to yield 
\be
\label{eq:ClBBapprox}
    C_\ell^{BB} \! \simeq \frac{36 \pi}{25} \! \left( \!\frac{\Delta \tau_r}{\tau_r} \! \!\right)^2 \! \!\int_0^\infty \!\frac{dk}{k} \mathcal{P}_h(k)\,j_2(k \tau_r)^2
    {\cal S}_\ell(k,\tau_0,0)^2,~~~~~~
\ee
where we have used the fact that modes entering during matter domination ($\tau > \tau_{\rm eq}$) obey 
\be
\mathcal{T}(\tau,k) \rightarrow \frac{  3 j_1(k \tau)}{k\tau}.
\ee
Note that the approximation in \Eq{eq:ClBBapprox} is inadequate to properly capture the behavior of $C_\ell^{BB}$ for large multipoles $\ell \gtrsim 100$. For this reason, we will use exclusively Eq.~(\ref{eq:ClBB}).

\section{Single and Double Bubble Spectra}\label{sec:bubblespectra}

Here we reproduce the explicit form for the function 
\be\label{eq:Delta}
\begin{split}
    \Delta\left( \frac{k}{\beta} \right) =  & \frac{16 G^2 \beta^2 k^3}{3 \kappa^2 H_*^4} \left( \frac{1+\alpha}{\alpha} \right)^2\\
 & \times \int_{t_i}^{t_f}  \!\!dt_1 \int_{t_i}^{t_f} \!\! dt_2  \cos[k(t_1 \! - \! t_2)] \,\Pi(t_1,t_2,k),~~~~~
\end{split}
\ee
originally derived in Ref.~\cite{Jinno:2016} working in the thin-wall and envelope approximations. As in this work, for simplicity we presume a luminal wall velocity $v_w = 1$, though the generalization to $v_w <1$ is straightforward. We also set $\beta = 1$ for convenience and restore it later as needed using dimensional analysis. We parametrize this function according to $\Delta = \Delta^{(1)} +  \Delta^{(2)}$, where the $\Delta^{(1),(2)}$ are the ``single-bubble'' and ``double-bubble'' contributions, respectively.

The ``single-bubble" term arises when the bubble wall segments passing through two distinct points originate from the same nucleation event, where 
\be
\label{eq:delta1}
     \Delta^{(1)} = \frac{k^3}{12\pi} \int_0^\infty dy  \int_{y}^\infty \frac{dr}{r^3} \, \frac{e^{-r/2} \cos(k y)}{\mathcal{I}(y, r)} {\cal{G}}(y, r, k),~~~~~~
\ee
where the integrand depends on the functions 
\be
    \mathcal{I}(y, r) &=& e^{y/r} + e^{-y/r} + \frac{y^2 - (r^2 + 4r)}{4r} e^{-r/2},~~~  \\
{\cal G}(y,r,k) &=& j_0(kr) F_0 + \frac{j_1(kr)}{kr} F_1 + \frac{j_2(kr)}{(k r)^2} F_2,~~
\ee
and we have defined 
\be
    F_0(y,r) &=& 2(r^2 - y^2)^2 (r^2 + 6r + 12)  \\ 
F_1(y,r) &=& 2(r^2 - y^2) [-r^2 (r^3 + 4r^2 + 12r + 24)   \nonumber  \\
    && + y^2 (r^3 + 12 r^2 + 60r + 120)]
   \\  
    F_2(y,r) &=& \frac{1}{2} \bigl[r^4 (r^4 + 4 r^3 + 20 r^2 + 72 r + 144)   \nonumber\\
    && - 2 y^2 r^2 (r^4 + 12 r^3 + 84 r^2 + 360 r + 720)  \nonumber \\
    && + y^4 (r^4 + 20 r^3 + 180 r^2 + 840 r + 1680)\bigr].~~~~~~~
\ee
The ``double-bubble'' contribution also
arises from bubble wall segments passing through  
two spatial points, but in this case the two wall segments originate from distinct nucleation events and 
\be
\label{eq:delta2}
    \Delta^{(2)} \!=\! \frac{k^3}{96\pi} \!  \int_0^{\infty} \!\!\! dy \!  \int_{y}^\infty
    \frac{ dr   \cos(k y) j_2(kr)
    g(y, r) g(- y, r) 
    }{ r^4 \mathcal{I}(y, r)^2 (k r)^2} ,~~ ~~~~ \,
\ee
where we have defined
\begin{equation}
    g(y, r) = (r^2 - y^2) [(r^3 + 2r^2) + y (r^2 + 6r + 12)] e^{-r/2} \,.
\end{equation}

\section{Source Term on Super-horizon Scales and Causality}\label{app.cauality}

As we have emphasized in the main text, phase transitions arise from causal processes occurring on sub-horizon scales. The fact that such a causal, sub-horizon source can excite modes that lie outside the horizon may at first glance seem to conflict with causality. Here, we demonstrate that having a non-vanishing source term ``on super-horizon scales'' does not violate causality. In real space, the source remains confined to sub-horizon regions and manifestly causal, in the sense that all real space correlations are constrained to vanish beyond the horizon. Nevertheless, analyticity in Fourier space ensures that power is also distributed among super-horizon modes, generically leading to a white noise power spectrum $\mathcal{P}(k) \propto k^3$ for $k/a \ll H$. 

Let $\tilde{\pi}_{ij}^T(t,\vec{x})$ be the real space transverse, traceless anisotropic stress tensor and let $\pi_{ij}^T(t,\vec{k})$ be its Fourier transform, defined as
\begin{equation}
    \pi_{ij}^T(t,\vec{k}) = \int d^3 x \, \tilde{\pi}_{ij}^T(t,\vec{x}) e^{i \vec{k} \cdot \vec{x}} \,.
\end{equation}
Let us further decompose $\pi_{ij}^T$ into two linearly independent polarizations
\begin{equation}
    \pi_{ij}^T(t,\vec{k}) = \sum_{\lambda = +, \times} \epsilon_{ij}^\lambda(\vec{k}) \pi_\lambda(t,k) \,.
\end{equation}
The \textit{dimensionful} power spectrum of the source $P_\pi$(k) is defined via the 2-point function as
\begin{equation}
    \langle \pi_\lambda(t,\vec{k}) \pi^*_{\lambda'}(t,\vec{k}) \rangle = (2\pi)^3 \delta^{(3)}(\vec{k} - \vec{k}') \delta_{\lambda \lambda'} P_\pi(t,\vec{k}) \,.
\end{equation}
We will also define the real space 2-point correlation function $\xi(r)$ as
\begin{equation}
    \langle \tilde{\pi}(t,\vec{x}) \tilde{\pi}^*(t,\vec{x}') \rangle \equiv \xi(r) \,,
\end{equation}
where $r = |\vec{x} - \vec{x}'|$. This is related to the power spectrum via
\begin{equation}\label{eq:Pkgendef}
    P_\pi(k) = \int d^3 r \, e^{i \vec{k} \cdot \vec{r}} \xi(r) = 4\pi \int_0^\infty dr \, r^2 \frac{\sin(kr)}{kr} \xi(r) \,.
\end{equation}

For a causal source, the real space correlation function will have compact support \textit{only} on scales less than the correlation scale $R$, which is necessarily subhorizon $R < H^{-1}$, i.e.
\begin{equation}\label{eq:xicaus}
    \xi(r) = \begin{cases} \xi(r) & r \leq R \,, \\ 0 & r > R \,. \end{cases}
\end{equation}
As a consequence, the Fourier transform will be analytic at $k = 0$, which generally implies that the spectrum will be ``white noise on large scales''. This is the Paley-Wiener theorem --- if a function $f(x)$ is compactly supported, then its Fourier transform $\tilde{f}(k)$ extends to an entire function of complex $k$. This means that it has a convergent power series everywhere in the complex plane, including $k=0$. This is quite easy to see. For the causal correlator in Eq.~(\ref{eq:xicaus}), Eq.~(\ref{eq:Pkgendef}) becomes
\begin{equation}\label{eq:Pkinprogress}
    P_\pi(k) = 4\pi \int_0^R dr \, r^2 \frac{\sin(kr)}{kr} \xi(r) \,.
\end{equation}
For small $k$, we expand the spherical Bessel function as $\frac{\sin(kr)}{kr} \simeq 1 - \frac{(kr)^2}{3!} + \frac{(kr)^4}{5!} - ...$ Inserting this into Eq.~(\ref{eq:Pkinprogress}), we find
\begin{equation}
    P_\pi(k) = A_0 + A_2 k^2 + A_4 k^4 + ...
\end{equation}
where we have defined the coefficients
\begin{equation}
    A_0 \! = \! 4\pi \int_0^R \!\!\!\! dr \, r^2 \xi(r),  \, A_2 \!=\! - \frac{4\pi}{3!} \int_0^R \!\!\!\! dr \, r^4 \xi(r), \, A_4 \!= \!...
\end{equation}
Now in terms of the \textit{dimensionless} power spectrum, $\mathcal{P}_\pi(k) = \frac{k^3}{2\pi^2} P_\pi(k)$, we see that the leading order scaling at small $k$ is
\begin{equation}
    \mathcal{P}_\pi(k) \sim k^3 \,.
\end{equation}
Thus we recover that the power spectrum of the source should decay on super-horizon scales as $k^3$. This white noise spectrum is an incredibly generic feature of causal sources and implies that fluctuations at sufficiently separated spatial points are uncorrelated.

In the context of bubble collisions sourced by a first order phase transition, the correlation scale will be set by the size of the largest bubbles at the end of the transition $R \sim v_w/\beta$, where $v_w$ is the bubble wall velocity and $\beta$ is the inverse time scale of the transition. The corresponding wavenumber $k_p \sim R^{-1}$ sets the peak of the spectrum. On larger scales (smaller $k$), we expect the spectrum to decay as white noise (i.e. $\sim k^3$) since these are length scales larger than the spatial correlations of our source. This intuition is indeed reflected in our results.

\section{Scalar perturbation bounds on cosmological phase transitions}\label{app.bound}

First-order phase transitions generate both tensor and scalar mode perturbations, resulting in curvature and isocurvature fluctuations constrained by observations such as the CMB, Lyman-$\alpha$, and CMB spectral distortion data.

\begin{figure}[t!]
\centering
\includegraphics[width=0.47\textwidth]{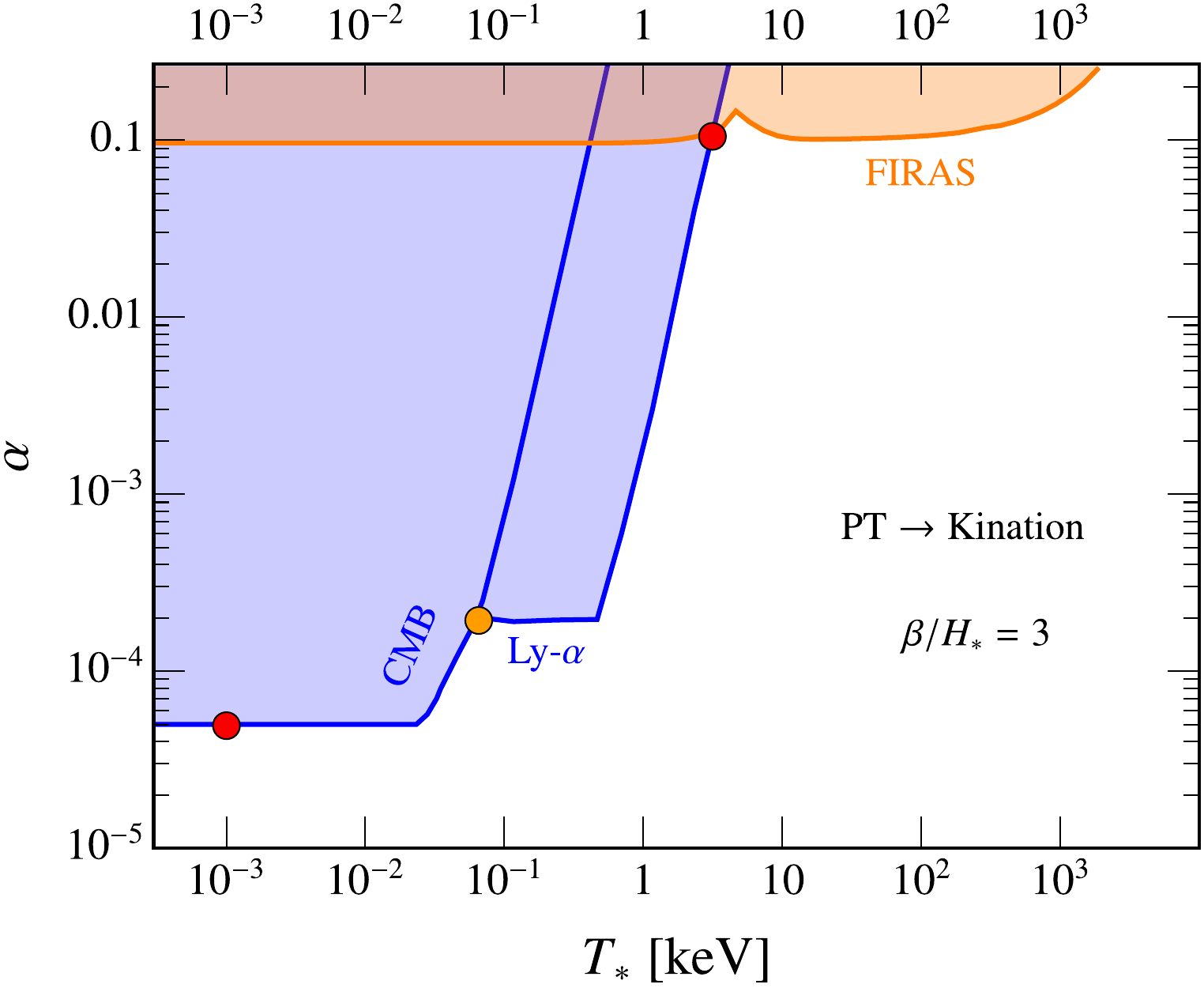}
\caption{Experimental limits ($2\sigma$) on $\alpha$ for various phase transition temperatures, assuming all energy is released into kination with the equation of state $w=1$. We show bounds from Lyman-$\alpha$ \cite{2011MNRAS.413.1717B} (blue shaded), CMB scalar perturbations \cite{Planck:2018jri} (blue shaded), and CMB spectral distortions \cite{Fixsen_2002} (orange shaded). The red circles denote the benchmark points of the main text; the upper-right point is $(T_*,\alpha) = (3\,{\rm keV},\,0.1)$, and the lower-left point is $(1\,{\rm eV},\,5\times 10^{-5})$. The orange circle at the intersection of the Lyman-$\alpha$ and CMB scalar perturbation bounds corresponds to $(71\, \text{eV}, 2\times10^{-4})$; we show the corresponding B-mode spectrum in Fig.~\ref{fig:bonusDlBB}. We assume $\kappa = 1$, $v_w=1$, and $\beta/H_* = 3$ in all cases. See Ref.~\cite{Elor:2023xbz} for a detailed derivation of the bounds.} 
\label{fig:bound}
\end{figure}

\begin{figure}[h!]
\centering
\includegraphics[width=0.5\textwidth]{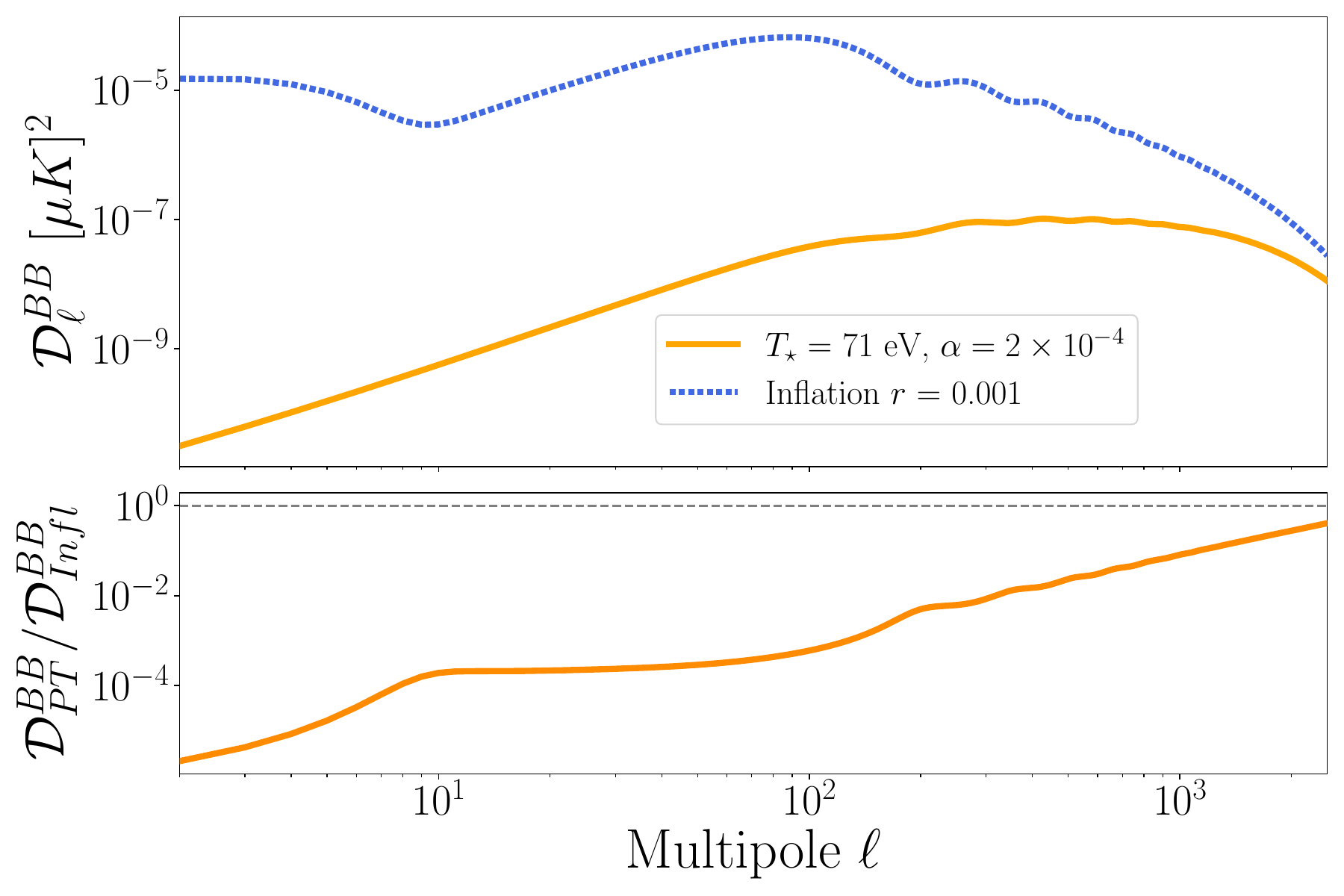}
\caption{\textbf{Top}: $B$-mode polarization spectra for the orange benchmark point of Fig.~\ref{fig:bound} and a minimal inflationary model with $r=0.001$ (dashed blue). 
{\bf Bottom}:
Ratio of $B$-mode signals from these sources.}
\label{fig:bonusDlBB}
\end{figure}

Several studies have examined scalar perturbation constraints from phase transitions, focusing on super-horizon perturbations~\cite{Liu:2022lvz, Elor:2023xbz,Buckley:2024nen} or sub-horizon perturbations in thermal phase transitions~\cite{Ramberg:2022irf}. Here, we review the findings of Ref.~\cite{Elor:2023xbz} for a hidden sector phase transition that converts latent heat into dark radiation and re-derive the constraints for a scenario where the phase transition energy is converted into kinetic energy which redshifts as $\rho \propto a^{-6}$, as in models of kination \cite{Gouttenoire:2021jhk}. The latter case significantly weakens the $\Delta N_{\rm eff}$ and scalar perturbation constraints, allowing for a $B$-mode signal potentially detectable by CMB-S4.

For a phase transition taking place at comoving time $\tau_{*}$, the super-horizon perturbation with comoving wavenumber $k$ contains $N\sim(k d_b)^3$ bubbles within the volume of the wavelength region, where $d_b=(8\pi)^{1/3}v_w/\beta$ is the average separation of bubbles. The phase transition time $t_c$ of each spatial point in the wavelength region has a standard deviation that goes as $1/\sqrt{N}$. Based on this, Ref.~\cite{Elor:2023xbz} shows that the super-horizon power spectrum of the phase transition time ${\cal P}_{\delta t} \equiv H_{*}^2\langle \delta t_c(\vec{x}) \delta t_c(\vec{y})\rangle$ has a parametric form \begin{equation}\label{eq:Pdt}
{\cal P}_{\delta t_c} = 8\pi c v_w^3 (k\tau_*)^3 \left(\frac{H_{*}}{\beta}\right)^5\,.
\end{equation}
A detailed calculation of the power spectrum performed in ~\cite{Elor:2023xbz}, which considers the bubble expansion dynamics, agrees with this parametric form and gives $c\approx 2.8$.

The curvature perturbation can be calculated in the spatially flat gauge by comparing energy densities at different spatial points at a common time when the scale factors are equal. Assuming an instantaneous transfer of the vacuum energy into dark fluid energy density $\rho_{\rm dark}$, the curvature perturbation given by the dark fluid is given by: 
\be
\label{eq:zeta_heu}
\hspace{-0.3cm}\zeta =-\frac{H_{\rm PT}\delta\rho_{\rm dark}}{(\dot{{\rho}}_{\rm dark}+\dot{\rho}_{\rm SM})}
=\frac{n\alpha H_{*}\delta t_c}{n\alpha+4(1-\alpha)\left(a_{\rm en}/a_{*}\right)^{n-4}}  ,~~ ~
\ee
where we have assumed the latent heat from the phase transition converts into energy density that redshifts as $\rho_{\rm dark}(a \geq a_*) = \rho_{\rm dark,ini} (a/a_*)^{-n}$, where $a_{\rm en}$ is the scale factor when the perturbation mode enters the horizon. We can write $(a_*/a_{\rm en}) = k \tau_*$ during the radiation-dominated era. For $n = 4$, corresponding to standard dark radiation, the dimensionless power spectrum from dark radiation reproduces the expression in~\cite{Elor:2023xbz} as  $P^{\rm DR}_\zeta = \alpha^2\mathcal{P}_{\delta t_c}$. 

In this work, we consider a phase transition example where the latent heat is converted into kination, which redshifts as $n = 6$. Since kination redshifts much faster than dark radiation, it is less affected by the $\Delta N_{\rm eff}$ bound. Moreover, its contribution to the total curvature perturbation is reduced due to the smaller energy ratio to the SM radiation just before the horizon entry. 

In the kination scenario, the total scalar perturbations right before horizon entry becomes 
\begin{equation}
\mathcal{P}_\zeta(k) = \frac{9}{4} \alpha^2(k\tau_*)^4\mathcal{P}_{\delta t_c}(k) + \mathcal{P}_{\rm ad}(k)\,.
\end{equation}
Here we have used the fact that $\alpha\ll1$ to simplify Eq.~(\ref{eq:zeta_heu}) when setting the exclusion bounds. Comparing the power spectrum to the 
CMB~\cite{Planck:2018jri}, Lyman-$\alpha$~\cite{2011MNRAS.413.1717B}, and CMB spectral distortion constraints~\cite{Fixsen_2002} on primordial curvature perturbations, one can obtain the exclusion bounds as the shaded regions in Fig.~\ref{fig:bound} for dark phase transition with $\beta/H_* = 2$. Our two benchmark models of the main text are indicated by red solid dots. 
For reference, we also show in Fig.~\ref{fig:bonusDlBB} the B-mode spectrum corresponding to the intersection of the Lyman-$\alpha$ and CMB scalar perturbation bounds (orange dot).

\bibliographystyle{utphys3}
\bibliography{biblio}

@article{Kite:2021,
    author = "Kite, Thomas and Chluba, Jens and Ravenni, Andrea and Patil, Subodh P.",
    title = "{Clarifying transfer function approximations for the large-scale gravitational wave background in \ensuremath{\Lambda}CDM}",
    eprint = "2107.13351",
    archivePrefix = "arXiv",
    primaryClass = "astro-ph.CO",
    doi = "10.1093/mnras/stab3125",
    journal = "Mon. Not. Roy. Astron. Soc.",
    volume = "509",
    number = "1",
    pages = "1366--1376",
    year = "2021"
}

@article{Geller:2021obo,
    author = "Geller, Michael and Lu, Sida and Tsai, Yuhsin",
    title = "{B modes from postinflationary gravitational waves sourced by axionic instabilities at cosmic reionization}",
    eprint = "2104.08284",
    archivePrefix = "arXiv",
    primaryClass = "hep-ph",
    doi = "10.1103/PhysRevD.104.083517",
    journal = "Phys. Rev. D",
    volume = "104",
    number = "8",
    pages = "083517",
    year = "2021"
}

@article{Kamionkowski:1993fg,
    author = "Kamionkowski, Marc and Kosowsky, Arthur and Turner, Michael S.",
    title = "{Gravitational radiation from first order phase transitions}",
    eprint = "astro-ph/9310044",
    archivePrefix = "arXiv",
    reportNumber = "IASSNS-HEP-93-44, FERMILAB-PUB-93-235-A",
    doi = "10.1103/PhysRevD.49.2837",
    journal = "Phys. Rev. D",
    volume = "49",
    pages = "2837--2851",
    year = "1994"
}

@article{Fixsen_2002,
doi = {10.1086/344402},
url = {https://dx.doi.org/10.1086/344402},
year = {2002},
month = {dec},
publisher = {},
volume = {581},
number = {2},
pages = {817},
author = {D. J. Fixsen and J. C. Mather},
title = {The Spectral Results of the Far-Infrared Absolute Spectrophotometer Instrument on COBE},
journal = {The Astrophysical Journal}
}

@ARTICLE{2011MNRAS.413.1717B,
       author = {{Bird}, Simeon and {Peiris}, Hiranya V. and {Viel}, Matteo and {Verde}, Licia},
        title = "{Minimally parametric power spectrum reconstruction from the Lyman {\ensuremath{\alpha}} forest}",
      journal = {MNRAS},
         year = 2011,
        month = may,
       volume = {413},
       number = {3},
        pages = {1717-1728},
          doi = {10.1111/j.1365-2966.2011.18245.x},
archivePrefix = {arXiv},
       eprint = {1010.1519},
 primaryClass = {astro-ph.CO},
       adsurl = {https://ui.adsabs.harvard.edu/abs/2011MNRAS.413.1717B}
}

@article{Planck:2018jri,
    author = "Akrami, Y. and others",
    collaboration = "Planck",
    title = "{Planck 2018 results. X. Constraints on inflation}",
    eprint = "1807.06211",
    archivePrefix = "arXiv",
    primaryClass = "astro-ph.CO",
    doi = "10.1051/0004-6361/201833887",
    journal = "Astron. Astrophys.",
    volume = "641",
    pages = "A10",
    year = "2020"
}

@article{Ramberg:2022irf,
    author = "Ramberg, Nicklas and Ratzinger, Wolfram and Schwaller, Pedro",
    title = "{One \ensuremath{\mu} to rule them all: CMB spectral distortions can probe domain walls, cosmic strings and low scale phase transitions}",
    eprint = "2209.14313",
    archivePrefix = "arXiv",
    primaryClass = "hep-ph",
    reportNumber = "MITP-22-077",
    doi = "10.1088/1475-7516/2023/02/039",
    journal = "JCAP",
    volume = "02",
    pages = "039",
    year = "2023"
}

@article{Liu:2022lvz,
    author = "Liu, Jing and Bian, Ligong and Cai, Rong-Gen and Guo, Zong-Kuan and Wang, Shao-Jiang",
    title = "{Constraining First-Order Phase Transitions with Curvature Perturbations}",
    eprint = "2208.14086",
    archivePrefix = "arXiv",
    primaryClass = "astro-ph.CO",
    doi = "10.1103/PhysRevLett.130.051001",
    journal = "Phys. Rev. Lett.",
    volume = "130",
    number = "5",
    pages = "051001",
    year = "2023"
}

@article{Elor:2023xbz,
    author = "Elor, Gilly and Jinno, Ryusuke and Kumar, Soubhik and McGehee, Robert and Tsai, Yuhsin",
    title = "{Finite Bubble Statistics Constrain Late Cosmological Phase Transitions}",
    eprint = "2311.16222",
    archivePrefix = "arXiv",
    primaryClass = "hep-ph",
    reportNumber = "FTPI-MINN-23-20, UTWI-39-2023",
    month = "11",
    year = "2023"
}

@article{BICEP:2021xfz,
    author = "Ade, P. A. R. and others",
    collaboration = "BICEP, Keck",
    title = "{Improved Constraints on Primordial Gravitational Waves using Planck, WMAP, and BICEP/Keck Observations through the 2018 Observing Season}",
    eprint = "2110.00483",
    archivePrefix = "arXiv",
    primaryClass = "astro-ph.CO",
    doi = "10.1103/PhysRevLett.127.151301",
    journal = "Phys. Rev. Lett.",
    volume = "127",
    number = "15",
    pages = "151301",
    year = "2021"
}

@article{CMB-S4:2016ple,
    author = "Abazajian, Kevork N. and others",
    collaboration = "CMB-S4",
    title = "{CMB-S4 Science Book, First Edition}",
    eprint = "1610.02743",
    archivePrefix = "arXiv",
    primaryClass = "astro-ph.CO",
    reportNumber = "FERMILAB-FN-1024-A-AE",
    month = "10",
    year = "2016"
}

@article{Kamionkowski:1996zd,
    author = "Kamionkowski, Marc and Kosowsky, Arthur and Stebbins, Albert",
    title = "{A Probe of primordial gravity waves and vorticity}",
    eprint = "astro-ph/9609132",
    archivePrefix = "arXiv",
    reportNumber = "CU-TP-767, CAL-615, FERMILAB-PUB-96-327-A",
    doi = "10.1103/PhysRevLett.78.2058",
    journal = "Phys. Rev. Lett.",
    volume = "78",
    pages = "2058--2061",
    year = "1997"
}

@article{Ellis:2023wic,
    author = "Ellis, John and Wands, David",
    title = "{Inflation (2023)}",
    eprint = "2312.13238",
    archivePrefix = "arXiv",
    primaryClass = "astro-ph.CO",
    reportNumber = "KCL-PH-TH/2023-73",
    month = "12",
    year = "2023"
}

@article{Caprini:2007,
    author = "Caprini, Chiara and Durrer, Ruth and Servant, Geraldine",
    title = "{Gravitational wave generation from bubble collisions in first-order phase transitions: An analytic approach}",
    eprint = "0711.2593",
    archivePrefix = "arXiv",
    primaryClass = "astro-ph",
    reportNumber = "CERN-PH-TH-2007-206, SACLAY-T07-142",
    doi = "10.1103/PhysRevD.77.124015",
    journal = "Phys. Rev. D",
    volume = "77",
    pages = "124015",
    year = "2008"
}

@article{Jinno:2016,
    author = "Jinno, Ryusuke and Takimoto, Masahiro",
    title = "{Gravitational waves from bubble collisions: An analytic derivation}",
    eprint = "1605.01403",
    archivePrefix = "arXiv",
    primaryClass = "astro-ph.CO",
    reportNumber = "KEK-TH-1900",
    doi = "10.1103/PhysRevD.95.024009",
    journal = "Phys. Rev. D",
    volume = "95",
    number = "2",
    pages = "024009",
    year = "2017"
}

@article{Hook:2020,
    author = "Hook, Anson and Marques-Tavares, Gustavo and Racco, Davide",
    title = "{Causal gravitational waves as a probe of free streaming particles and the expansion of the Universe}",
    eprint = "2010.03568",
    archivePrefix = "arXiv",
    primaryClass = "hep-ph",
    doi = "10.1007/JHEP02(2021)117",
    journal = "JHEP",
    volume = "02",
    pages = "117",
    year = "2021"
}

@article{Caprini:2018,
    author = "Caprini, Chiara and Figueroa, Daniel G.",
    title = "{Cosmological Backgrounds of Gravitational Waves}",
    eprint = "1801.04268",
    archivePrefix = "arXiv",
    primaryClass = "astro-ph.CO",
    doi = "10.1088/1361-6382/aac608",
    journal = "Class. Quant. Grav.",
    volume = "35",
    number = "16",
    pages = "163001",
    year = "2018"
}

@article{Buckley:2024nen,
    author = "Buckley, Matthew R. and Du, Peizhi and Fernandez, Nicolas and Weikert, Mitchell J.",
    title = "{Dark radiation isocurvature from cosmological phase transitions}",
    eprint = "2402.13309",
    archivePrefix = "arXiv",
    primaryClass = "hep-ph",
    doi = "10.1088/1475-7516/2024/07/031",
    journal = "JCAP",
    volume = "07",
    pages = "031",
    year = "2024"
}

@article{Lesgourgues:2011,
    author = "Lesgourgues, Julien",
    title = "{The Cosmic Linear Anisotropy Solving System (CLASS) I: Overview}",
    eprint = "1104.2932",
    archivePrefix = "arXiv",
    primaryClass = "astro-ph.IM",
    month = "4",
    year = "2011"
}

@article{Blas:2011,
    author = "Blas, Diego and Lesgourgues, Julien and Tram, Thomas",
    title = "{The Cosmic Linear Anisotropy Solving System (CLASS) II: Approximation schemes}",
    eprint = "1104.2933",
    archivePrefix = "arXiv",
    primaryClass = "astro-ph.CO",
    reportNumber = "CERN-PH-TH-2011-082, LAPTH-010-11",
    doi = "10.1088/1475-7516/2011/07/034",
    journal = "JCAP",
    volume = "07",
    pages = "034",
    year = "2011"
}

@book{Rubakov:2017,
    author = "Rubakov, Valery A. and Gorbunov, Dmitry S.",
    title = "{Introduction to the Theory of the Early Universe}: {Hot big bang theory}",
    doi = "10.1142/10447",
    isbn = "978-981-320-987-9, 978-981-320-988-6, 978-981-322-005-8",
    publisher = "World Scientific",
    address = "Singapore",
    year = "2017"
}

@article{Kosowsky:1994,
    author = "Kosowsky, Arthur",
    title = "{Cosmic microwave background polarization}",
    eprint = "astro-ph/9501045",
    archivePrefix = "arXiv",
    reportNumber = "FERMILAB-PUB-94-206-A",
    doi = "10.1006/aphy.1996.0020",
    journal = "Annals Phys.",
    volume = "246",
    pages = "49--85",
    year = "1996"
}

@article{Lasky:2015lej,
    author = "Lasky, Paul D. and others",
    title = "{Gravitational-wave cosmology across 29 decades in frequency}",
    eprint = "1511.05994",
    archivePrefix = "arXiv",
    primaryClass = "astro-ph.CO",
    doi = "10.1103/PhysRevX.6.011035",
    journal = "Phys. Rev. X",
    volume = "6",
    number = "1",
    pages = "011035",
    year = "2016"
}

@article{Kamionkowski:2015yta,
    author = "Kamionkowski, Marc and Kovetz, Ely D.",
    title = "{The Quest for B Modes from Inflationary Gravitational Waves}",
    eprint = "1510.06042",
    archivePrefix = "arXiv",
    primaryClass = "astro-ph.CO",
    doi = "10.1146/annurev-astro-081915-023433",
    journal = "Ann. Rev. Astron. Astrophys.",
    volume = "54",
    pages = "227--269",
    year = "2016"
}

@article{Seljak:1996gy,
    author = "Seljak, Uros and Zaldarriaga, Matias",
    title = "{Signature of gravity waves in polarization of the microwave background}",
    eprint = "astro-ph/9609169",
    archivePrefix = "arXiv",
    doi = "10.1103/PhysRevLett.78.2054",
    journal = "Phys. Rev. Lett.",
    volume = "78",
    pages = "2054--2057",
    year = "1997"
}

@article{Kamionkowski:1996ks,
    author = "Kamionkowski, Marc and Kosowsky, Arthur and Stebbins, Albert",
    title = "{Statistics of cosmic microwave background polarization}",
    eprint = "astro-ph/9611125",
    archivePrefix = "arXiv",
    reportNumber = "FERMILAB-PUB-96-426-A, CU-TP-787, CAL-617",
    doi = "10.1103/PhysRevD.55.7368",
    journal = "Phys. Rev. D",
    volume = "55",
    pages = "7368--7388",
    year = "1997"
}

@article{Zaldarriaga:1996xe,
    author = "Zaldarriaga, Matias and Seljak, Uros",
    title = "{An all sky analysis of polarization in the microwave background}",
    eprint = "astro-ph/9609170",
    archivePrefix = "arXiv",
    doi = "10.1103/PhysRevD.55.1830",
    journal = "Phys. Rev. D",
    volume = "55",
    pages = "1830--1840",
    year = "1997"
}

@article{Seljak:1996ti,
    author = "Seljak, Uros",
    title = "{Measuring polarization in cosmic microwave background}",
    eprint = "astro-ph/9608131",
    archivePrefix = "arXiv",
    reportNumber = "CFA-96-8",
    doi = "10.1086/304123",
    journal = "Astrophys. J.",
    volume = "482",
    pages = "6",
    year = "1997"
}

@article{Guzzetti:2016mkm,
    author = "Guzzetti, M. C. and Bartolo, N. and Liguori, M. and Matarrese, S.",
    title = "{Gravitational waves from inflation}",
    eprint = "1605.01615",
    archivePrefix = "arXiv",
    primaryClass = "astro-ph.CO",
    doi = "10.1393/ncr/i2016-10127-1",
    journal = "Riv. Nuovo Cim.",
    volume = "39",
    number = "9",
    pages = "399--495",
    year = "2016"
}

@article{Dev:2016feu,
    author = "Dev, P. S. Bhupal and Mazumdar, A.",
    title = "{Probing the Scale of New Physics by Advanced LIGO/VIRGO}",
    eprint = "1602.04203",
    archivePrefix = "arXiv",
    primaryClass = "hep-ph",
    doi = "10.1103/PhysRevD.93.104001",
    journal = "Phys. Rev. D",
    volume = "93",
    number = "10",
    pages = "104001",
    year = "2016"
}

@article{Caprini:2015zlo,
    author = "Caprini, Chiara and others",
    title = "{Science with the space-based interferometer eLISA. II: Gravitational waves from cosmological phase transitions}",
    eprint = "1512.06239",
    archivePrefix = "arXiv",
    primaryClass = "astro-ph.CO",
    reportNumber = "DESY-15-246",
    doi = "10.1088/1475-7516/2016/04/001",
    journal = "JCAP",
    volume = "04",
    pages = "001",
    year = "2016"
}

@article{Weir:2017wfa,
    author = "Weir, David J.",
    title = "{Gravitational waves from a first order electroweak phase transition: a brief review}",
    eprint = "1705.01783",
    archivePrefix = "arXiv",
    primaryClass = "hep-ph",
    reportNumber = "HIP-2017-06-TH, HIP-2017-06/TH",
    doi = "10.1098/rsta.2017.0126",
    journal = "Phil. Trans. Roy. Soc. Lond. A",
    volume = "376",
    number = "2114",
    pages = "20170126",
    year = "2018",
    note = "[Erratum: Phil.Trans.Roy.Soc.Lond.A 381, 20230212 (2023)]"
}

@article{Hindmarsh:2020hop,
    author = {Hindmarsh, Mark B. and L\"uben, Marvin and Lumma, Johannes and Pauly, Martin},
    title = "{Phase transitions in the early universe}",
    eprint = "2008.09136",
    archivePrefix = "arXiv",
    primaryClass = "astro-ph.CO",
    reportNumber = "MPP-2020-163, HIP-2020-27/TH",
    doi = "10.21468/SciPostPhysLectNotes.24",
    journal = "SciPost Phys. Lect. Notes",
    volume = "24",
    pages = "1",
    year = "2021"
}

@article{Kosowsky:1992rz,
    author = "Kosowsky, Arthur and Turner, Michael S. and Watkins, Richard",
    title = "{Gravitational waves from first order cosmological phase transitions}",
    reportNumber = "FERMILAB-PUB-91-333-A-REV, FERMILAB-PUB-91-333-A",
    doi = "10.1103/PhysRevLett.69.2026",
    journal = "Phys. Rev. Lett.",
    volume = "69",
    pages = "2026--2029",
    year = "1992"
}

@article{Ellis:2020awk,
    author = "Ellis, John and Lewicki, Marek and No, Jos\'e Miguel",
    title = "{Gravitational waves from first-order cosmological phase transitions: lifetime of the sound wave source}",
    eprint = "2003.07360",
    archivePrefix = "arXiv",
    primaryClass = "hep-ph",
    reportNumber = "KCL-PH-TH/2020-04, CERN-TH-2020-016, IFT-UAM/CSIC-20-35",
    doi = "10.1088/1475-7516/2020/07/050",
    journal = "JCAP",
    volume = "07",
    pages = "050",
    year = "2020"
}

@article{Cutting:2020nla,
    author = "Cutting, Daniel and Escartin, Elba Granados and Hindmarsh, Mark and Weir, David J.",
    title = "{Gravitational waves from vacuum first order phase transitions II: from thin to thick walls}",
    eprint = "2005.13537",
    archivePrefix = "arXiv",
    primaryClass = "astro-ph.CO",
    reportNumber = "HIP-2020-13/TH",
    doi = "10.1103/PhysRevD.103.023531",
    journal = "Phys. Rev. D",
    volume = "103",
    number = "2",
    pages = "023531",
    year = "2021"
}

@inproceedings{Baumann:2009ds,
    author = "Baumann, Daniel",
    title = "{Inflation}",
    booktitle = "{Theoretical Advanced Study Institute in Elementary Particle Physics}: {Physics of the Large and the Small}",
    eprint = "0907.5424",
    archivePrefix = "arXiv",
    primaryClass = "hep-th",
    reportNumber = "TASI-2009",
    doi = "10.1142/9789814327183_0010",
    pages = "523--686",
    year = "2011"
}

@article{Hawking:1982ga,
    author = "Hawking, S. W. and Moss, I. G. and Stewart, J. M.",
    title = "{Bubble Collisions in the Very Early Universe}",
    reportNumber = "Print-82-0180 (CAMBRIDGE)",
    doi = "10.1103/PhysRevD.26.2681",
    journal = "Phys. Rev. D",
    volume = "26",
    pages = "2681",
    year = "1982"
}

@article{Espinosa:2010hh,
    author = "Espinosa, Jose R. and Konstandin, Thomas and No, Jose M. and Servant, Geraldine",
    title = "{Energy Budget of Cosmological First-order Phase Transitions}",
    eprint = "1004.4187",
    archivePrefix = "arXiv",
    primaryClass = "hep-ph",
    reportNumber = "CERN-PH-TH-2010-027",
    doi = "10.1088/1475-7516/2010/06/028",
    journal = "JCAP",
    volume = "06",
    pages = "028",
    year = "2010"
}

@article{Huber:2008hg,
    author = "Huber, Stephan J. and Konstandin, Thomas",
    title = "{Gravitational Wave Production by Collisions: More Bubbles}",
    eprint = "0806.1828",
    archivePrefix = "arXiv",
    primaryClass = "hep-ph",
    doi = "10.1088/1475-7516/2008/09/022",
    journal = "JCAP",
    volume = "09",
    pages = "022",
    year = "2008"
}

@article{Hindmarsh:2015qta,
    author = "Hindmarsh, Mark and Huber, Stephan J. and Rummukainen, Kari and Weir, David J.",
    title = "{Numerical simulations of acoustically generated gravitational waves at a first order phase transition}",
    eprint = "1504.03291",
    archivePrefix = "arXiv",
    primaryClass = "astro-ph.CO",
    reportNumber = "HIP-2015-13-TH",
    doi = "10.1103/PhysRevD.92.123009",
    journal = "Phys. Rev. D",
    volume = "92",
    number = "12",
    pages = "123009",
    year = "2015"
}

@article{Hindmarsh:2013xza,
    author = "Hindmarsh, Mark and Huber, Stephan J. and Rummukainen, Kari and Weir, David J.",
    title = "{Gravitational waves from the sound of a first order phase transition}",
    eprint = "1304.2433",
    archivePrefix = "arXiv",
    primaryClass = "hep-ph",
    reportNumber = "HIP-2013-07-TH",
    doi = "10.1103/PhysRevLett.112.041301",
    journal = "Phys. Rev. Lett.",
    volume = "112",
    pages = "041301",
    year = "2014"
}

@article{Grojean:2006bp,
    author = "Grojean, Christophe and Servant, Geraldine",
    title = "{Gravitational Waves from Phase Transitions at the Electroweak Scale and Beyond}",
    eprint = "hep-ph/0607107",
    archivePrefix = "arXiv",
    reportNumber = "CERN-PH-TH-2006-125",
    doi = "10.1103/PhysRevD.75.043507",
    journal = "Phys. Rev. D",
    volume = "75",
    pages = "043507",
    year = "2007"
}

@article{Hindmarsh:2017gnf,
    author = "Hindmarsh, Mark and Huber, Stephan J. and Rummukainen, Kari and Weir, David J.",
    title = "{Shape of the acoustic gravitational wave power spectrum from a first order phase transition}",
    eprint = "1704.05871",
    archivePrefix = "arXiv",
    primaryClass = "astro-ph.CO",
    reportNumber = "HIP-2017-02-TH, HIP-2017-02/TH",
    doi = "10.1103/PhysRevD.96.103520",
    journal = "Phys. Rev. D",
    volume = "96",
    number = "10",
    pages = "103520",
    year = "2017",
    note = "[Erratum: Phys.Rev.D 101, 089902 (2020)]"
}

@article{Schwaller:2015tja,
    author = "Schwaller, Pedro",
    title = "{Gravitational Waves from a Dark Phase Transition}",
    eprint = "1504.07263",
    archivePrefix = "arXiv",
    primaryClass = "hep-ph",
    reportNumber = "CERN-PH-TH-2015-093",
    doi = "10.1103/PhysRevLett.115.181101",
    journal = "Phys. Rev. Lett.",
    volume = "115",
    number = "18",
    pages = "181101",
    year = "2015"
}

@article{Caprini:2009fx,
    author = "Caprini, Chiara and Durrer, Ruth and Konstandin, Thomas and Servant, Geraldine",
    title = "{General Properties of the Gravitational Wave Spectrum from Phase Transitions}",
    eprint = "0901.1661",
    archivePrefix = "arXiv",
    primaryClass = "astro-ph.CO",
    doi = "10.1103/PhysRevD.79.083519",
    journal = "Phys. Rev. D",
    volume = "79",
    pages = "083519",
    year = "2009"
}

@article{Kosowsky:2001xp,
    author = "Kosowsky, Arthur and Mack, Andrew and Kahniashvili, Tinatin",
    title = "{Gravitational radiation from cosmological turbulence}",
    eprint = "astro-ph/0111483",
    archivePrefix = "arXiv",
    reportNumber = "RAP-334",
    doi = "10.1103/PhysRevD.66.024030",
    journal = "Phys. Rev. D",
    volume = "66",
    pages = "024030",
    year = "2002"
}

@article{Apreda:2001us,
    author = "Apreda, Riccardo and Maggiore, Michele and Nicolis, Alberto and Riotto, Antonio",
    title = "{Gravitational waves from electroweak phase transitions}",
    eprint = "gr-qc/0107033",
    archivePrefix = "arXiv",
    reportNumber = "UGVA-DPT-07-1096",
    doi = "10.1016/S0550-3213(02)00264-X",
    journal = "Nucl. Phys. B",
    volume = "631",
    pages = "342--368",
    year = "2002"
}

@article{Cutting:2018tjt,
    author = "Cutting, Daniel and Hindmarsh, Mark and Weir, David J.",
    title = "{Gravitational waves from vacuum first-order phase transitions: from the envelope to the lattice}",
    eprint = "1802.05712",
    archivePrefix = "arXiv",
    primaryClass = "astro-ph.CO",
    reportNumber = "HIP-2018-4-TH",
    doi = "10.1103/PhysRevD.97.123513",
    journal = "Phys. Rev. D",
    volume = "97",
    number = "12",
    pages = "123513",
    year = "2018"
}

@article{Hindmarsh:2019phv,
    author = "Hindmarsh, Mark and Hijazi, Mulham",
    title = "{Gravitational waves from first order cosmological phase transitions in the Sound Shell Model}",
    eprint = "1909.10040",
    archivePrefix = "arXiv",
    primaryClass = "astro-ph.CO",
    reportNumber = "NORDITA-2019-083, HIP-2019-29/TH",
    doi = "10.1088/1475-7516/2019/12/062",
    journal = "JCAP",
    volume = "12",
    pages = "062",
    year = "2019"
}

@article{Jinno:2017fby,
    author = "Jinno, Ryusuke and Takimoto, Masahiro",
    title = "{Gravitational waves from bubble dynamics: Beyond the Envelope}",
    eprint = "1707.03111",
    archivePrefix = "arXiv",
    primaryClass = "hep-ph",
    reportNumber = "CTPU-17-26, KEK-TH-1986",
    doi = "10.1088/1475-7516/2019/01/060",
    journal = "JCAP",
    volume = "01",
    pages = "060",
    year = "2019"
}

@article{Brdar:2018num,
    author = "Brdar, Vedran and Helmboldt, Alexander J. and Kubo, Jisuke",
    title = "{Gravitational Waves from First-Order Phase Transitions: LIGO as a Window to Unexplored Seesaw Scales}",
    eprint = "1810.12306",
    archivePrefix = "arXiv",
    primaryClass = "hep-ph",
    doi = "10.1088/1475-7516/2019/02/021",
    journal = "JCAP",
    volume = "02",
    pages = "021",
    year = "2019"
}

@article{Gouttenoire:2021jhk,
    author = "Gouttenoire, Yann and Servant, Geraldine and Simakachorn, Peera",
    title = "{Kination cosmology from scalar fields and gravitational-wave signatures}",
    eprint = "2111.01150",
    archivePrefix = "arXiv",
    primaryClass = "hep-ph",
    reportNumber = "DESY 21-134",
    month = "11",
    year = "2021"
}

@article{Kahniashvili:2008pf,
    author = "Kahniashvili, Tina and Kosowsky, Arthur and Gogoberidze, Grigol and Maravin, Yurii",
    title = "{Detectability of Gravitational Waves from Phase Transitions}",
    eprint = "0806.0293",
    archivePrefix = "arXiv",
    primaryClass = "astro-ph",
    doi = "10.1103/PhysRevD.78.043003",
    journal = "Phys. Rev. D",
    volume = "78",
    pages = "043003",
    year = "2008"
}

@article{Athron:2023xlk,
    author = "Athron, Peter and Bal\'azs, Csaba and Fowlie, Andrew and Morris, Lachlan and Wu, Lei",
    title = "{Cosmological phase transitions: From perturbative particle physics to gravitational waves}",
    eprint = "2305.02357",
    archivePrefix = "arXiv",
    primaryClass = "hep-ph",
    doi = "10.1016/j.ppnp.2023.104094",
    journal = "Prog. Part. Nucl. Phys.",
    volume = "135",
    pages = "104094",
    year = "2024"
}

@article{Nicolis:2003tg,
    author = "Nicolis, Alberto",
    title = "{Relic gravitational waves from colliding bubbles and cosmic turbulence}",
    eprint = "gr-qc/0303084",
    archivePrefix = "arXiv",
    reportNumber = "IEM-FT-230-03",
    doi = "10.1088/0264-9381/21/4/L05",
    journal = "Class. Quant. Grav.",
    volume = "21",
    pages = "L27",
    year = "2004"
}

@article{Seljak:2003pn,
    author = "Seljak, Uros and Hirata, Christopher M.",
    title = "{Gravitational lensing as a contaminant of the gravity wave signal in CMB}",
    eprint = "astro-ph/0310163",
    archivePrefix = "arXiv",
    doi = "10.1103/PhysRevD.69.043005",
    journal = "Phys. Rev. D",
    volume = "69",
    pages = "043005",
    year = "2004"
}

@article{Kogut:2024vbi,
    author = "Kogut, Alan and others",
    title = "{The Primordial Inflation Explorer (PIXIE): Mission Design and Science Goals}",
    eprint = "2405.20403",
    archivePrefix = "arXiv",
    primaryClass = "astro-ph.CO",
    month = "5",
    year = "2024"
}

@article{Sabyr:2024lgg,
    author = "Sabyr, Alina and Sierra, Carlos and Hill, J. Colin and McMahon, Jeffrey J.",
    title = "{SPECTER: An Instrument Concept for CMB Spectral Distortion Measurements with Enhanced Sensitivity}",
    eprint = "2409.12188",
    archivePrefix = "arXiv",
    primaryClass = "astro-ph.CO",
    month = "9",
    year = "2024"
}

@article{Mortonson:2009qv,
    author = "Mortonson, Michael J. and Dvorkin, Cora and Peiris, Hiranya V. and Hu, Wayne",
    title = "{CMB polarization features from inflation versus reionization}",
    eprint = "0903.4920",
    archivePrefix = "arXiv",
    primaryClass = "astro-ph.CO",
    doi = "10.1103/PhysRevD.79.103519",
    journal = "Phys. Rev. D",
    volume = "79",
    pages = "103519",
    year = "2009"
}

@article{Dodelson:2003bv,
    author = "Dodelson, Scott and Rozo, Eduardo and Stebbins, Albert",
    title = "{Primordial gravity waves and weak lensing}",
    eprint = "astro-ph/0301177",
    archivePrefix = "arXiv",
    reportNumber = "FERMILAB-PUB-03-011-A",
    doi = "10.1103/PhysRevLett.91.021301",
    journal = "Phys. Rev. Lett.",
    volume = "91",
    pages = "021301",
    year = "2003"
}

@article{Bernardeau:1996aa,
    author = "Bernardeau, F.",
    title = "{Weak lensing detection in CMB maps}",
    eprint = "astro-ph/9611012",
    archivePrefix = "arXiv",
    reportNumber = "SACLAY-SPH-T-96-122",
    journal = "Astron. Astrophys.",
    volume = "324",
    pages = "15--26",
    year = "1997"
}

@article{Hanson:2009kr,
    author = "Hanson, Duncan and Challinor, Anthony and Lewis, Antony",
    title = "{Weak lensing of the CMB}",
    eprint = "0911.0612",
    archivePrefix = "arXiv",
    primaryClass = "astro-ph.CO",
    doi = "10.1007/s10714-010-1036-y",
    journal = "Gen. Rel. Grav.",
    volume = "42",
    pages = "2197--2218",
    year = "2010"
}

@article{Zaldarriaga:1998ar,
    author = "Zaldarriaga, Matias and Seljak, Uros",
    title = "{Gravitational lensing effect on cosmic microwave background polarization}",
    eprint = "astro-ph/9803150",
    archivePrefix = "arXiv",
    doi = "10.1103/PhysRevD.58.023003",
    journal = "Phys. Rev. D",
    volume = "58",
    pages = "023003",
    year = "1998"
}

@article{Hu:2001kj,
    author = "Hu, Wayne and Okamoto, Takemi",
    title = "{Mass reconstruction with cmb polarization}",
    eprint = "astro-ph/0111606",
    archivePrefix = "arXiv",
    doi = "10.1086/341110",
    journal = "Astrophys. J.",
    volume = "574",
    pages = "566--574",
    year = "2002"
}

@article{Hu:2001fb,
    author = "Hu, Wayne",
    title = "{Dark synergy: Gravitational lensing and the CMB}",
    eprint = "astro-ph/0108090",
    archivePrefix = "arXiv",
    doi = "10.1103/PhysRevD.65.023003",
    journal = "Phys. Rev. D",
    volume = "65",
    pages = "023003",
    year = "2002"
}

@article{Hu:2002vu,
    author = "Hu, Wayne and Hedman, Matthew M. and Zaldarriaga, Matias",
    title = "{Benchmark parameters for CMB polarization experiments}",
    eprint = "astro-ph/0210096",
    archivePrefix = "arXiv",
    doi = "10.1103/PhysRevD.67.043004",
    journal = "Phys. Rev. D",
    volume = "67",
    pages = "043004",
    year = "2003"
}

@article{Seljak:1995ve,
    author = "Seljak, Uros",
    title = "{Gravitational lensing effect on cosmic microwave background anisotropies: A Power spectrum approach}",
    eprint = "astro-ph/9505109",
    archivePrefix = "arXiv",
    reportNumber = "MIT-CSR-94-29",
    doi = "10.1086/177218",
    journal = "Astrophys. J.",
    volume = "463",
    pages = "1",
    year = "1996"
}

@article{Dodelson:2010qu,
    author = "Dodelson, Scott",
    title = "{Cross-Correlating Probes of Primordial Gravitational Waves}",
    eprint = "1001.5012",
    archivePrefix = "arXiv",
    primaryClass = "astro-ph.CO",
    reportNumber = "FERMILAB-PUB-10-020-A-PPD-T",
    doi = "10.1103/PhysRevD.82.023522",
    journal = "Phys. Rev. D",
    volume = "82",
    pages = "023522",
    year = "2010"
}

@article{Hu:1997hv,
    author = "Hu, Wayne and White, Martin J.",
    title = "{A CMB polarization primer}",
    eprint = "astro-ph/9706147",
    archivePrefix = "arXiv",
    doi = "10.1016/S1384-1076(97)00022-5",
    journal = "New Astron.",
    volume = "2",
    pages = "323",
    year = "1997"
}

@article{Trendafilova:2023xtq,
    author = "Trendafilova, Cynthia and Hotinli, Selim C. and Meyers, Joel",
    title = "{Improving constraints on inflation with CMB delensing}",
    eprint = "2312.02954",
    archivePrefix = "arXiv",
    primaryClass = "astro-ph.CO",
    doi = "10.1088/1475-7516/2024/06/017",
    journal = "JCAP",
    volume = "06",
    pages = "017",
    year = "2024"
}

@article{Hotinli:2021umk,
    author = "Hotinli, Selim C. and Meyers, Joel and Trendafilova, Cynthia and Green, Daniel and van Engelen, Alexander",
    title = "{The benefits of CMB delensing}",
    eprint = "2111.15036",
    archivePrefix = "arXiv",
    primaryClass = "astro-ph.CO",
    doi = "10.1088/1475-7516/2022/04/020",
    journal = "JCAP",
    volume = "04",
    number = "04",
    pages = "020",
    year = "2022"
}

@article{Lewis:2006fu,
    author = "Lewis, Antony and Challinor, Anthony",
    title = "{Weak gravitational lensing of the CMB}",
    eprint = "astro-ph/0601594",
    archivePrefix = "arXiv",
    doi = "10.1016/j.physrep.2006.03.002",
    journal = "Phys. Rept.",
    volume = "429",
    pages = "1--65",
    year = "2006"
}

@article{Geller:2018mwu,
    author = "Geller, Michael and Hook, Anson and Sundrum, Raman and Tsai, Yuhsin",
    title = "{Primordial Anisotropies in the Gravitational Wave Background from Cosmological Phase Transitions}",
    eprint = "1803.10780",
    archivePrefix = "arXiv",
    primaryClass = "hep-ph",
    doi = "10.1103/PhysRevLett.121.201303",
    journal = "Phys. Rev. Lett.",
    volume = "121",
    number = "20",
    pages = "201303",
    year = "2018"
}

@article{Smith:2005mm,
    author = "Smith, Tristan L. and Kamionkowski, Marc and Cooray, Asantha",
    title = "{Direct detection of the inflationary gravitational wave background}",
    eprint = "astro-ph/0506422",
    archivePrefix = "arXiv",
    doi = "10.1103/PhysRevD.73.023504",
    journal = "Phys. Rev. D",
    volume = "73",
    pages = "023504",
    year = "2006"
}

@article{Alanne:2019bsm,
    author = "Alanne, Tommi and Hugle, Thomas and Platscher, Moritz and Schmitz, Kai",
    title = "{A fresh look at the gravitational-wave signal from cosmological phase transitions}",
    eprint = "1909.11356",
    archivePrefix = "arXiv",
    primaryClass = "hep-ph",
    reportNumber = "CERN-TH-2019-158",
    doi = "10.1007/JHEP03(2020)004",
    journal = "JHEP",
    volume = "03",
    pages = "004",
    year = "2020"
}

@article{Schmitz:2020syl,
    author = "Schmitz, Kai",
    title = "{New Sensitivity Curves for Gravitational-Wave Signals from Cosmological Phase Transitions}",
    eprint = "2002.04615",
    archivePrefix = "arXiv",
    primaryClass = "hep-ph",
    reportNumber = "CERN-TH-2020-018",
    doi = "10.1007/JHEP01(2021)097",
    journal = "JHEP",
    volume = "01",
    pages = "097",
    year = "2021"
}

@article{Caprini:2018mtu,
    author = "Caprini, Chiara and Figueroa, Daniel G.",
    title = "{Cosmological Backgrounds of Gravitational Waves}",
    eprint = "1801.04268",
    archivePrefix = "arXiv",
    primaryClass = "astro-ph.CO",
    doi = "10.1088/1361-6382/aac608",
    journal = "Class. Quant. Grav.",
    volume = "35",
    number = "16",
    pages = "163001",
    year = "2018"
}

@article{Mazumdar:2018dfl,
    author = "Mazumdar, Anupam and White, Graham",
    title = "{Review of cosmic phase transitions: their significance and experimental signatures}",
    eprint = "1811.01948",
    archivePrefix = "arXiv",
    primaryClass = "hep-ph",
    doi = "10.1088/1361-6633/ab1f55",
    journal = "Rept. Prog. Phys.",
    volume = "82",
    number = "7",
    pages = "076901",
    year = "2019"
}

@article{Kosowsky:1992vn,
    author = "Kosowsky, Arthur and Turner, Michael S.",
    title = "{Gravitational radiation from colliding vacuum bubbles: envelope approximation to many bubble collisions}",
    eprint = "astro-ph/9211004",
    archivePrefix = "arXiv",
    reportNumber = "FERMILAB-PUB-92-295-A",
    doi = "10.1103/PhysRevD.47.4372",
    journal = "Phys. Rev. D",
    volume = "47",
    pages = "4372--4391",
    year = "1993"
}

@article{Hindmarsh:2016lnk,
    author = "Hindmarsh, Mark",
    title = "{Sound shell model for acoustic gravitational wave production at a first-order phase transition in the early Universe}",
    eprint = "1608.04735",
    archivePrefix = "arXiv",
    primaryClass = "astro-ph.CO",
    doi = "10.1103/PhysRevLett.120.071301",
    journal = "Phys. Rev. Lett.",
    volume = "120",
    number = "7",
    pages = "071301",
    year = "2018"
}

@article{Caprini:2006jb,
    author = "Caprini, Chiara and Durrer, Ruth",
    title = "{Gravitational waves from stochastic relativistic sources: Primordial turbulence and magnetic fields}",
    eprint = "astro-ph/0603476",
    archivePrefix = "arXiv",
    doi = "10.1103/PhysRevD.74.063521",
    journal = "Phys. Rev. D",
    volume = "74",
    pages = "063521",
    year = "2006"
}

@article{Kahniashvili:2008pe,
    author = "Kahniashvili, Tina and Campanelli, Leonardo and Gogoberidze, Grigol and Maravin, Yurii and Ratra, Bharat",
    title = "{Gravitational Radiation from Primordial Helical Inverse Cascade MHD Turbulence}",
    eprint = "0809.1899",
    archivePrefix = "arXiv",
    primaryClass = "astro-ph",
    doi = "10.1103/PhysRevD.78.123006",
    journal = "Phys. Rev. D",
    volume = "78",
    pages = "123006",
    year = "2008",
    note = "[Erratum: Phys.Rev.D 79, 109901 (2009)]"
}

@article{Caprini:2009yp,
    author = "Caprini, Chiara and Durrer, Ruth and Servant, Geraldine",
    title = "{The stochastic gravitational wave background from turbulence and magnetic fields generated by a first-order phase transition}",
    eprint = "0909.0622",
    archivePrefix = "arXiv",
    primaryClass = "astro-ph.CO",
    doi = "10.1088/1475-7516/2009/12/024",
    journal = "JCAP",
    volume = "12",
    pages = "024",
    year = "2009"
}

@article{Kahniashvili:2009mf,
    author = "Kahniashvili, Tina and Kisslinger, Leonard and Stevens, Trevor",
    title = "{Gravitational Radiation Generated by Magnetic Fields in Cosmological Phase Transitions}",
    eprint = "0905.0643",
    archivePrefix = "arXiv",
    primaryClass = "astro-ph.CO",
    doi = "10.1103/PhysRevD.81.023004",
    journal = "Phys. Rev. D",
    volume = "81",
    pages = "023004",
    year = "2010"
}

@article{Kisslinger:2015hua,
    author = "Kisslinger, Leonard and Kahniashvili, Tina",
    title = "{Polarized Gravitational Waves from Cosmological Phase Transitions}",
    eprint = "1505.03680",
    archivePrefix = "arXiv",
    primaryClass = "astro-ph.CO",
    doi = "10.1103/PhysRevD.92.043006",
    journal = "Phys. Rev. D",
    volume = "92",
    number = "4",
    pages = "043006",
    year = "2015"
}

@article{Caprini:2024gyk,
    author = "Caprini, Chiara and Jinno, Ryusuke and Konstandin, Thomas and Roper Pol, Alberto and Rubira, Henrique and Stomberg, Isak",
    title = "{Gravitational waves from decaying sources in strong phase transitions}",
    eprint = "2409.03651",
    archivePrefix = "arXiv",
    primaryClass = "gr-qc",
    month = "9",
    year = "2024"
}

@article{RoperPol:2023dzg,
    author = "Roper Pol, Alberto and Procacci, Simona and Caprini, Chiara",
    title = "{Characterization of the gravitational wave spectrum from sound waves within the sound shell model}",
    eprint = "2308.12943",
    archivePrefix = "arXiv",
    primaryClass = "gr-qc",
    doi = "10.1103/PhysRevD.109.063531",
    journal = "Phys. Rev. D",
    volume = "109",
    number = "6",
    pages = "063531",
    year = "2024"
}

@article{Auclair:2022jod,
    author = "Auclair, Pierre and Caprini, Chiara and Cutting, Daniel and Hindmarsh, Mark and Rummukainen, Kari and Steer, Dani\`ele A. and Weir, David J.",
    title = "{Generation of gravitational waves from freely decaying turbulence}",
    eprint = "2205.02588",
    archivePrefix = "arXiv",
    primaryClass = "astro-ph.CO",
    reportNumber = "HIP-2021-35/TH",
    doi = "10.1088/1475-7516/2022/09/029",
    journal = "JCAP",
    volume = "09",
    pages = "029",
    year = "2022"
}

@article{Caprini:2015tfa,
    author = "Caprini, Chiara",
    editor = "Ciani, Giacomo and Conklin, John W. and Mueller, Guido",
    title = "{Stochastic background of gravitational waves from cosmological sources}",
    eprint = "1501.01174",
    archivePrefix = "arXiv",
    primaryClass = "gr-qc",
    doi = "10.1088/1742-6596/610/1/012004",
    journal = "J. Phys. Conf. Ser.",
    volume = "610",
    number = "1",
    pages = "012004",
    year = "2015"
}

@article{Weinberg:2003ur,
    author = "Weinberg, Steven",
    title = "{Damping of tensor modes in cosmology}",
    eprint = "astro-ph/0306304",
    archivePrefix = "arXiv",
    reportNumber = "UTTG-02-03",
    doi = "10.1103/PhysRevD.69.023503",
    journal = "Phys. Rev. D",
    volume = "69",
    pages = "023503",
    year = "2004"
}

@article{Flauger:2007es,
    author = "Flauger, Raphael and Weinberg, Steven",
    title = "{Tensor Microwave Background Fluctuations for Large Multipole Order}",
    eprint = "astro-ph/0703179",
    archivePrefix = "arXiv",
    reportNumber = "UTTG-01-07",
    doi = "10.1103/PhysRevD.75.123505",
    journal = "Phys. Rev. D",
    volume = "75",
    pages = "123505",
    year = "2007"
}

@article{Baumann:2014cja,
    author = "Baumann, Daniel and Green, Daniel and Porto, Rafael A.",
    title = "{B-modes and the Nature of Inflation}",
    eprint = "1407.2621",
    archivePrefix = "arXiv",
    primaryClass = "hep-th",
    doi = "10.1088/1475-7516/2015/01/016",
    journal = "JCAP",
    volume = "01",
    pages = "016",
    year = "2015"
}

@article{BICEP2:2014owc,
    author = "Ade, P. A. R. and others",
    collaboration = "BICEP2",
    title = "{Detection of $B$-Mode Polarization at Degree Angular Scales by BICEP2}",
    eprint = "1403.3985",
    archivePrefix = "arXiv",
    primaryClass = "astro-ph.CO",
    doi = "10.1103/PhysRevLett.112.241101",
    journal = "Phys. Rev. Lett.",
    volume = "112",
    number = "24",
    pages = "241101",
    year = "2014"
}

@article{Janssen:2014dka,
    author = "Janssen, Gemma and others",
    editor = "Bourke, Tyler L. and others",
    title = "{Gravitational wave astronomy with the SKA}",
    eprint = "1501.00127",
    archivePrefix = "arXiv",
    primaryClass = "astro-ph.IM",
    doi = "10.22323/1.215.0037",
    journal = "PoS",
    volume = "AASKA14",
    pages = "037",
    year = "2015"
}

@article{LISA:2017pwj,
    author = "Amaro-Seoane, Pau and others",
    collaboration = "LISA",
    title = "{Laser Interferometer Space Antenna}",
    eprint = "1702.00786",
    archivePrefix = "arXiv",
    primaryClass = "astro-ph.IM",
    month = "2",
    year = "2017"
}

@article{Cai:2019,
    author = "Cai, Rong-Gen and Pi, Shi and Sasaki, Misao",
    title = "{Universal infrared scaling of gravitational wave background spectra}",
    eprint = "1909.13728",
    archivePrefix = "arXiv",
    primaryClass = "astro-ph.CO",
    reportNumber = "IPMU19-0135, YITP-19-88",
    doi = "10.1103/PhysRevD.102.083528",
    journal = "Phys. Rev. D",
    volume = "102",
    number = "8",
    pages = "083528",
    year = "2020"
}

@article{Ellis:2019,
    author = "Ellis, John and Lewicki, Marek and No, Jos\'e Miguel and Vaskonen, Ville",
    title = "{Gravitational wave energy budget in strongly supercooled phase transitions}",
    eprint = "1903.09642",
    archivePrefix = "arXiv",
    primaryClass = "hep-ph",
    reportNumber = "KCL-PH-TH/2019-32, CERN-TH-2019-032, IFT-UAM/CSIC-19-32",
    doi = "10.1088/1475-7516/2019/06/024",
    journal = "JCAP",
    volume = "06",
    pages = "024",
    year = "2019"
}

@article{Yamada:2025long,
    author = "Yamada, Masaki",
    title = "{Analytic derivation of GW spectrum from bubble collisions in FLRW Universe}",
    eprint = "2509.16073",
    archivePrefix = "arXiv",
    primaryClass = "astro-ph.CO",
    reportNumber = "TU-1276",
    month = "9",
    year = "2025"
}

@article{Yamada:2025short,
    author = "Yamada, Masaki",
    title = "{Maximal GW amplitude from bubble collisions in supercooled phase transitions}",
    eprint = "2509.13402",
    archivePrefix = "arXiv",
    primaryClass = "gr-qc",
    reportNumber = "TU-1275",
    month = "9",
    year = "2025"
}

\end{document}